\documentclass{elsarticle}

\journal{Physica A}

\usepackage{amsmath}    
\usepackage{amsfonts}  
\usepackage{graphicx}   
\usepackage{bm}  

\usepackage{calligra}
\DeclareMathAlphabet{\mathcalligra}{T1}{calligra}{m}{n}
\DeclareFontShape{T1}{calligra}{m}{n}{<->s*[2.2]callig15}{}

\newcommand{\scripty}[1]{\ensuremath{\mathcalligra{#1}}}
\def\scrmag{\scripty{r}}

\def\scrhat{\hat{\scripty{r}}}
\def\scr2{\frac{\scrhat}{\scrmag^2}}
\def\rvec{\overrightarrow{r}}

\def\del{\overrightarrow{\nabla}}

\def\Fvec{\overrightarrow{F}}

\def\Evec{\overrightarrow{E}}

\def\jvec{\overrightarrow{j}}

\begin{document}
\begin{frontmatter}
\title{Thermodynamically Induced Particle Transport: Order-by-Induction and Entropic Trapping at the Nano-Scale} 

\author{S. N. Patitsas}
\ead{steve.patitsas@uleth.ca}
\address{University of Lethbridge,\\4401 University Drive, Lethbridge AB, Canada, T1K3M4}



\date{\today}

\begin{abstract}

A theory for thermodynamic induction (TI) under isothermal conditions is presented.  This includes a treatment of the Helmholtz free energy budget available for a gate variable to utilize towards aiding another variable's approach towards thermodynamic equilibrium.  This energy budget could be used to help create interesting physical structures and examples of order-by-induction.    I also show how to treat TI in the continuum limit which can be obtained from a variational principle. 
 
Several important examples of isothermal TI have been discussed, including a type of electromigration that may be detectable in electrolytes, superfluids and semiconductors. As an example of a bottlenecked system exhibiting enhanced TI, manipulation of atoms and molecules by STM has been discussed in detail.  My considerations provide strong support for microscopic bond-breaking mechanisms being governed by a general thermodynamic principle.  In particular, I show that induced entropy trapping can explain the level of control that sliding-type manipulations demonstrate.  The most reasonable choices for the parameters input into the simple formula give a threshold condition for STM manipulations that is strikingly close to what is required to match results reported in the literature.  My continuum model predicts the shape of the adsorbate potential well for the STM case and from this I predict a level of force detectable by AFM.  A final proposal, and example of order-by-induction, predict a long tether may be constructed between sample and tip that is just one atom thick.
\end{abstract}
\end{frontmatter}


\maketitle 

\section{Introduction} \label{sec:intro}

Recently, the Onsager relations \cite{Onsager1931,Onsager1931b} have been extended from the linear to the nonlinear realm inside the class of nonequilibrium thermodynamic systems governed by variable kinetic coefficients (VKC)~\cite{Patitsas2014}.  Though the result is not quite as simple to express as $L_{ik}=L_{ki}$ is, the extended result may be expressed concisely as $N_{ki}=-r N_{ik}$ where $r$ is a ratio of two relevant time scales.  
That kinetic, or transport, coefficients are not always constant is well known~\cite{Miller1960,Prigogine1965,mazur} and the direct $N_{ik}$ coefficients arise as 1st order corrections to kinetic coefficients.  The rate of the relaxation process for state variable $x_i$ depends on the gating variable $x_k$.   The main result of Ref.~\cite{Patitsas2014} is the existence of the previously unknown $N_{ki}$ coefficients (across the diagonal).  These induction coefficients change the dynamics of gating variables.  For example, such variables can be induced to move away from equilibrium by a significant amount.  Thermodynamic induction (TI) may be described as the adjustment of the gate variable in order to enhance the approach towards equilibrium of variable $x_i$, the dynamical reservoir (DR).    

In Ref.~\cite{Patitsas2014} it was suggested that systems that have a bottlenecking or gating property might be good candidates for displaying measurable TI effects.  In addition TI predicts the enhancement of transport rates for a bottlenecked process, as a consequence of maximizing entropy production~\cite{Patitsas2014}.   Entropy production also plays a key role in studies of bottlenecked systems exhibiting so-called entropic barriers, which are distinguished from traditional energy barriers~\cite{Zwanzig1992a,Zwanzig1992b,PhysRevE.64.061106,PhysRevLett.96.130603}.  An interesting result of transport process dominated by entropy is the entropy trap which can be used to trap macromolecules under nonequilibrium conditions~\cite{Asher1999,Craighead1999}.  The entropy trap works because the system fluctuates and attempts to sample all available microstates.
In this work I explore the possibility of entropy traps being created as a consequence of TI.  This will require coupling of at least two transport processes, i.e., coupling between a DR and a gate, or control variable.
One area of activity that warrants careful inspection for TI and any resulting entropic trapping effects is the highly controlled manipulation of atoms and molecules by STM.  Scanning tunneling microscopy is an inherently nonequilibrium technique because of the applied bias across the tunnel junction.  It is well known that adsorbates underneath the tip have strong effects on the junction conductance.  Thus it is reasonable to explore the possibility that adsorbates near the junction might be affected by TI effects.

The STM technique has recently been established as a powerful tool that can be widely used to consistently modify surfaces in such ways.  Basic manipulation mechanisms that have been clearly established may be grouped into the following categories: a) Near-contact type, b) Electronic excitation of adsorbates, c) Ladder-type multiple vibrational excitation, and d) Single electron resonant excitation.  Examples of a) include the lateral pushing/sliding of Xe atoms on Pt(111) at low temperatures, as well as the reversible transfer of Xe between tip and sample, where in both cases the tip is placed very close to the surface~\cite{Eigler91}. An example of b) is H desorption from Si(100) at $>6$~V applied bias~\cite{AvourisLyding96}.  Examples of c) include the dissociation of O$_2$ molecules~\cite{HoPersson97} and desorption of hydrogen from silicon surfaces~\cite{AvourisPersson97}, while examples of d) include the reversible transfer of CO between tip and sample~\cite{Meyer1998} and the desorption of organic molecules from silicon~\cite{Patitsas2000,PatitsasPRL2000,Polanyi1999,Palmer2005}.

Subsequent work has introduced many refinements: For example, several of the basic mechanisms have been employed to carry out the full Ullmann reaction on two specific iodobenzene molecules~\cite{Meyer2000}.  Also, large molecules can now be rotated and translated by small amounts on a surface, TBPP for example~\cite{Meyer2001}.  Atomic manipulation can even be used as a characterization tool for identifying products of surface dissociation studies, for example chlorine atoms from trichloroethylene~\cite{Patitsas2007}.  Also noteworthy are recent results showing non-local STM induced manipulations (See \cite{Palmer2010} and references within) where it appears that the spread of injected carriers over $\approx$ 100 \AA~ after tunneling plays an important role.  

I begin, in Sec.~\ref{sec:gentheory}, by developing a theory for thermodynamic induction involving $n$ variables, under isothermal conditions, and deriving dynamical equations for the approach to equilibrium.  The isothermal approach is naturally based on the Helmholtz free energy and is well-suited for describing particle transport.  In Sec.~\ref{statstates} the importance of the quasistationary states is pointed out in the form of an important inequality regarding rates of free energy production.  This is followed, in Sec.~\ref{sec:maxent}, by the establishment of a variational principle governing the minimization of a free power function specifically designed for the VKC systems.  Important two-variable example systems are discussed in Sec.~\ref{sec:2var} and Sec.~\ref{sec:examples}.  
Before concluding, I demonstrate in Sec.~\ref{sec:cont} how TI is incorporated into the limit of continuous variables. This section is a significant step towards incorporating TI into more complex continuum systems such as diffusing systems and systems with fluid flow.

\section{General Theory for Isothermal Thermodynamic Induction}  \label{sec:gentheory}

I consider a system A described by $n$ thermodynamic variables $x_i$ with equilibrium values $x_{i_0}$.  In this work I follow closely the approach presented in Ref.~\cite{Patitsas2014}, with the main difference being that in this treatment I assume isothermal conditions, i.e., $dT=0$ throughout.  This approach involves a coupling of system A to a large heat bath, or reservoir, always held at temperature $T$.  This traditional sort of reservoir will play a passive role in the analysis and is not to be confused with the DR which I introduced in Ref.~\cite{Patitsas2014} and plays a prominent role in my analysis below.  I begin by considering only discrete variables, and subsequently develop a continuum theory in Sec.~\ref{sec:cont}.    

For system A, the Helmholtz free energy, $F=U-TS$, may be written as a function of the $n$ variables $x_i$.  The heat bath is not included, i.e., all subsystems considered below are part of system A, not the heat bath.  The positive definite change $\Delta F$, from equilibrium, of the total free energy, may be written as a function of the $n$ state variables 
$a_i=x_i-x_{i_0}$.
\begin{equation}
\Delta F\equiv F(x_1,x_2,...,x_n)-F(x_{1_0},x_{2_0},...,x_{n_0})= \frac{T}{2}\sum_{pq} c_{pq}^{-1}a_p a_q~.    \label{quadF}
\end{equation}
Generalized forces (affinities) are defined by	
\begin{equation}
Y_p= -\frac{1}{T}\frac{\partial \Delta F}{\partial a_p}~.\label{Yp}
\end{equation}
These conjugate parameters are interrelated as:   
\begin{equation}
a_p=-\sum_{q=1}^n c_{pq}Y_q       ~  ,    \label{agX}
\end{equation}  
and
\begin{equation}
Y_p=-\sum_{q=1}^n c_{pq}^{-1} a_q         ~,    \label{Xga}
\end{equation} 
where $c_{pq}=c_{qp}$, i.e., the generalized capacitance matrix is symmetric, and with the same symmetry holding for the inverse matrix.  In order to ensure thermodynamic stability, both the capacitance matrix and its inverse must be positive definite.  These relations place the $Y_p$ variables on equal footing with the $a_p$ variables.

When considering relaxation towards equilibrium, the customary approach is to write dynamical equations that relate the time derivatives $\dot{a}_i$ to linear combinations of the forces $Y_j$.  The kinetic coefficients are closely related to transport coefficients such as thermal conductivity, electrical conductivity, etc.  These transport processes are irreversible:  the total system entropy increases with time as the system approaches equilibrium. Likewise, the Helmholtz free energy $F$ for system A always decreases with time as system A approaches equilibrium.  In terms of the rate of free energy production, $\dot{F}=\frac{dF}{dt}$, I note that
\begin{equation}
\dot{F}= -T\sum_{p=1}^n Y_p \dot{a}_p    ~.  \label{Fdot}
\end{equation}
This bilinear form plays an important role in the upcoming analysis by allowing a general way to express $\Delta F$ and then by extracting the nonlinear contribution to $\Delta F$.

\subsection{Dynamics}

In this approach, a non-equilibrium average value for a variable is determined by evaluating an equilibrium average accompanied by the insertion of a $\exp(-\Delta F/k_B T)$ probability weighting factor. Note that it must be the total free energy change for system A that is used for this purpose.  Explicitly then for variable $a_p$:  
\begin{equation}
\langle \dot{a}_p(t')\rangle=\left\langle \dot{a}_p(t')e^{-\beta\Delta F(t'-t)}\right\rangle_0~,   \label{expS}          
\end{equation} 
where $\beta=1/(k_B T)$, and $\Delta F(t'-t)\equiv F(t')-F(t)$. The brackets $\langle~~\rangle_0$ denote ensemble averaging over equilibrium states.  Since the change in free energy is small, a linear expansion is warranted, leaving:
\begin{equation}
\langle \dot{a}_p(t')\rangle= -\beta{\langle \dot{a}_p(t')\Delta F(t'-t)}\rangle_0~,   \label{adot2b}
\end{equation}
since $\langle \dot{a}_i \rangle_0=0$.  It is worth noting that the results similar to Eqs.~(\ref{expS},\ref{adot2b}) can be established by utilizing a perturbative quantum mechanical approach reviewed by Bernard and Callen, and then taking the high temperature classical limit~\cite{Callen1959}.  This approach also implements a bilinear form involving variable and conjugate force, as the perturbation term.  In the classical limit, one finds this bilinear form as the argument of an exponential function of a Boltzmann-factor, very similarly to Eq.~(\ref{expS}).  This theory has been successfully applied to the general study of the relationships between fluctuations and response functions, to first order, second order, and beyond.  Integrating both sides of Eq.~(\ref{adot2b}) over the time interval $\Delta t_i$ gives the coarse-grained time derivative:
\begin{equation}
\bar{\dot{a}}_{i}=\frac{1}{\Delta t_i}\int_t^{t+\Delta t_i}\langle\dot{a}_i(t')\rangle dt'=\frac{-\beta}{\Delta t_i}\int_{t}^{t+\Delta t_i}{dt'\langle\dot{a}_i(t')\Delta F(t'-t)\rangle_0}~. \label{adot3}
\end{equation} 
A bar is used to denote the coarse-graining and it is understood that the time step $\Delta t_i$ is much larger that the correlation time $\tau_i^*$ for the random force driving the fluctuations.

\subsection{Review of linear case}

Substituting Eq.~(\ref{DeltaFint}) into Eq.~(\ref{adot3}) leaves
\begin{equation}
\bar{\dot{a}}_i= \frac{1}{k_B \Delta t_i}\int_{t}^{t+\Delta t_i}{dt' \int_{t}^{t+\Delta t_i}dt''\langle\dot{a}_i(t')\sum_{j=1}^n Y_j \dot{a}_j(t'')\rangle_0}~. 
\end{equation}
Closely following Ref.~\cite{Reif15} by taking the slowly varying force functions $Y_j$ out of the integrals, and with a few more elementary steps one obtains
\begin{equation}
\bar{\dot{a}}_i=\sum_{j=1}^n L_{ij}Y_j     \label{Lij} ~,
\end{equation}
where 
\begin{equation}
L_{ij}=\frac{1}{k_B}\int_{-\infty}^0 ds K_{ij}(s)~,  \label{LijKij}
\end{equation}
and $K_{ij}$ are the cross-correlation functions defined by $K_{ij}\equiv\langle \dot{a}_i(t)\dot{a}_j(t+s)\rangle_0$.  The Onsager reciprocal symmetry $L_{ij}=L_{ji}$ is obtained by using time-reversal symmetry and assumption of zero external magnetic field~\cite{Reif15}.  In the linear regime, the kinetic coefficients $L_{ij}$ are constants.  One actually solves for the system dynamics by combining Eqs.~(\ref{agX},~\ref{Lij}) to obtain
\begin{equation}
{\bar{\dot{Y}}}_i=-\sum_{j=1}^n A_{ij}Y_j     \label{Xlindyneq} ~,
\end{equation}
where $A_{ij}=\sum_{k=1}^n c_{ik}^{-1} L_{kj}$.  The eigenvalues of the $A$ matrix have units of $s^{-1}$ and these define the $n$ (relaxation) timescales $\tau_i$ for the system.  In the special case where the matrices $c_{kl}$ and $A_{kl}$ are diagonal, then I have the following relation:
\begin{equation}
L_{kk}\tau_k= c_{kk}     \label{gLtauOne} ~.
\end{equation}
After combining Eq.~(\ref{Fdot}) and Eq.~(\ref{Lij}), I determine the rate of total free energy production to be 
\begin{equation}
\bar{\dot{F}} = -T\sum_{p=1}^n \sum_{q=1}^n L_{pq}Y_p Y_q   ~.               \label{dotF}
\end{equation} 
Finally, for the linear case I may express the coarse-grained $\Delta F$ as
\begin{equation}
\Delta F=\int_{t}^{t+\Delta t_i}{\bar{\dot{F}} dt''}=-T\sum_{p=1}^n \int_{t}^{t+\Delta t_i}{ Y_j \bar{\dot{a}}_j dt''}=-T\sum_{p=1}^n\sum_{q=1}^n \int_{t}^{t+\Delta t_i}{L_{pq}Y_p Y_q dt''}    ~ \label{DeltaFint}
\end{equation}
where I made use of Eq.~(\ref{Fdot}).

\subsection{Nonlinear dynamics}

Next I consider the case where the kinetic coefficients are not constants but instead depend on the variables $\{x_i\}$, i.e., the case of VKC systems.  It is here that the key result for TI is derived, i.e., Eq.~(\ref{relateIndDir}).  My aim then is to adapt the basic approach used in the linear analysis by adding in nonlinear terms.  The new (variable) coefficients will be labeled $M_{ij}(\{a_l\})$.  In equilibrium, where $a_l=0$, they take the (constant) values $L_{ij}$, i.e., $M_{ij}(\{a_l=0\})=L_{ij}$.
The important assumptions made that allow derivation of the induction effect have been discussed in detail in Ref.~\cite{Patitsas2014}. The same assumptions are made here.  Briefly, I make linear expansions in the variables $\{a_l\}$, and ignore higher order terms.  This defines a new set of constants $\gamma_{ij,l}$ such that
\begin{equation}
M_{ij}=L_{ij}+\sum_{l}\gamma_{ij,l} a_l~,       \label{gamma}               
\end{equation}
and
\begin{equation}
\gamma_{ij,l}=\left(\frac{\partial M_{ij}}{\partial a_l}\right)_{a_l=0}   ~.\label{gammaijk}
\end{equation}
The $\gamma_{ij,l}$ coefficients, which have the symmetry, $\gamma_{ij,l}=\gamma_{ji,l}$, describe the variability of the kinetic coefficients to leading order.  The next important assumption is that of the $n$ variables in question, $m$ are of the slow variety while $n-m$ are quickly varying in time.  The slow variables are labeled with indices $i,j$ with index values ranging from 1 to $m$, and the fast variables are labeled with indices $k,l$ with these index values ranging from $m+1$ to $n$.   More exactly the assumption is $\tau_k<<\tau_i$ for all $k>m$ and $i\leq m$.  
This distinction between slow and fast variables is not new and has been described in detail elsewhere in the context of simplifying the dynamical equations for large systems by eliminating the fast variables~\cite{VanKampen1985}.  Here, the approach is similar and useful equations for slow variable dynamics will be obtained, though I emphasize that I am also interested in the fast variable dynamics. 
I also assume no variability in the kinetic coefficients $M_{kl}$ associated with fast variables i.e. $\gamma_{kl,q}=0$ for all $k,l>m$, 
and that the $M_{ij}$ coefficients (for the slow variables) depend only on fast variables i.e. $\gamma_{ij,q}=0$ for $q\leq m$.   

The dynamical equations are modified to become
\begin{equation}
\bar{\dot{a}}_p=\sum_{q=1}^n (L_{pq}+N_{pq})Y_q  ~.   \label{Neqs} 
\end{equation}
Nonlinear effects are manifested in the slowly-varying $N_{pq}$ coefficients which are derived in a procedure very similar to the proof of the principle of thermodynamic induction~\cite{Patitsas2014}.   
The key step in the derivation is to revisit Eq.~(\ref{adot3}) and make a distinction between linear and nonlinear contributions to $\Delta F$, i.e., $\Delta F=\Delta F_{lin}+\Delta F_{nonlin}$.
Use of Eq.~(\ref{dotF}) with the substitution $L_{ij}\rightarrow M_{ij}$ (also using  Eq.~(\ref{gamma})) and extracting the nonlinear terms gives 
\begin{equation}
\bar{\dot{F}}_{nonlin}=-T\sum_{i=1}^{m}\sum_{j=1}^{m}\sum_{l=m+1}^{n}\gamma_{ij,l} Y_i Y_j a_l ~.    \label{FdotNonlin}
\end{equation}  
Substituting Eq.~(\ref{FdotNonlin}) into Eq.~(\ref{adot3}) gives the desired result in steps very similar to those presented in Ref.~\cite{Patitsas2014}:
\begin{equation}
\left(\bar{\dot{a}}_k\right)_{ind}=\frac{1}{k_B\Delta t_k}\sum_{i=1}^{m}Y_i\sum_{j=1}^{m}Y_j\sum_{l=m+1}^{n}\gamma_{ji,l}\int_{t}^{t+\Delta t_k}dt'\int_t^{t'}dt'' \int_{-\infty}^{t''}dt'''\langle\dot{a}_k(t') \dot{a}_l(t''') \rangle_0~.    \label{akindfull}
\end{equation}
Equation~(\ref{akindfull}) is suitable for taking the continuum limit in Section~\ref{sec:cont} below.
The induction terms in Eq.~(\ref{Neqs}) are obtained by noting that $\langle\dot{a}_k(t') \dot{a}_l(t''')\rangle_0= K_{kl}(t'-t''')$ is the correlation function and by making use of Eq.~(\ref{LijKij}), i.e., that the fast block part of the $M$ matrix is assumed diagonal.  See Ref.~\cite{Patitsas2014} for the derivation of the direct terms.
The direct and induced terms are simply related as
\begin{equation}
N_{ki}=-r_k N_{ik} ~, \label{relateIndDir}
\end{equation}
where the dimensionless ratio $r_k$ is defined as 
\begin{equation}
r_k\equiv \frac{\tau_k^*}{\tau_k} ~,
\end{equation}
and 
\begin{equation}
N_{ki}= L_{kk} \tau^*_{k} \sum_{j=1}^{m}\gamma_{ij,k} Y_j ~,~~~~k>m ,~i\leq m~.  \label{Nind}
\end{equation}
Equation~(\ref{relateIndDir}) is the most direct mathematical representation for the TI effect, here under isothermal conditions.  It is remarkable that $N_{ki}\neq 0$.

For the fast variables $a_k$
\begin{equation}
\bar{\dot{a}}_k=\sum_{p=1}^n (L_{kp}+N_{kp})Y_p  ~.   \label{Neqsfast} 
\end{equation}
In these equations the $N_{kl}$ terms (on lower left of matrix) are not a direct result of the variability of kinetic coefficients, but are instead a result of thermodynamic induction.  Thus, the fast variables are induced to change in ways that are not obvious.  For example, even if the fast subsystems begin in thermodynamic equilibrium, they may be pushed out of equilibrium, at least while some slow variables remain out of equilibrium. 
The induced terms are of opposite sign than the corresponding direct terms.  They are also smaller in magnitude since I have assumed $r_k<<1$.  

This concludes the general discussion and leads to special cases and possible physical examples.  Note that a clear distinction has been made between the so-called slow variables, which I also call DR variables, and fast variables, which I also call gate, or control, variables. A DR is an out of equilibrium variable, with a virtually limitless supply that approaches equilibrium very slowly.  An example is the charge in an RC circuit with a very large capacitor.  If the conductance of a DR depends on the particular mean value of a gate variable, then these gate variables control the rate at which the DR approaches equilibrium.

For convenience, I drop the coarse-graining bar symbol below.

\section{Quasistationary states} \label{statstates}

The concept of stationary states, as described in Refs.~\cite{mazur,Prigogine} in the linear regime, is still applicable in this nonlinear context.  For this discussion, I feel it necessary to make a subtle distinction between (completely) stationary states and quasistationary states.  For quasistationary conditions, the variables $a_i$ evolve slowly in time.  If I formally take the limit where all of the slow timescales $\tau_i$ go to infinity, then I have the conditions for stationary states to exist.    
For a physical example, an electrical circuit involving a capacitor slowly draining charge (with large RC time constant) corresponds to quasistationary conditions.  Replacing the capacitor with a power supply will hold the applied bias indefinitely over time and would create stationary conditions.  Quasistationary states have actually been studied previously~\cite{VanKampen1985,Poston1981}, though with different terminology used, such as ''adiabatic elimination'', ''entrainment'', and ''slaving''.  In the general case as treated by van Kampen, after some transient motion, fast variables settle, or become entrained, into a slowly varying state that ''tracks'' the slow variable.  An example is the current acting as the entrained fast variable in a highly overdamped electrical circuit with a slowly varying applied potential.  

Following Ref.~\cite{mazur} and defining the fluxes $J_p\equiv \bar{\dot{a}}_p$, quasistationary states are defined by the conditions: $J_k=0$, for $k>m$.  Explicitly: 
\begin{equation}
J_k= L_{kk}Y_k+ \sum_{i=1}^{m} N_{ki}Y_i =0   ~~.    \label{Jk}  
\end{equation}
Solving and using Eq.~(\ref{Nind}) gives:
\begin{equation}
Y_k|_{qss}=- \frac{1}{ L_{kk}}\sum_{i=1}^{m} N_{ki}Y_i = - \tau^*_{k}\sum_{i=1}^{m} \sum_{j=1}^{m}\gamma_{ij,k}   Y_i Y_j  ~.  \label{XfastStat}    
\end{equation}

When a given fast variable $x_k$ is quasistationary, it is out of equilibrium, and that subsystem free energy is raised from the equilibrium value by a net amount $\Delta F_k$ given by
\begin{equation}
\Delta F_{k}= -\frac{T}{2}Y_k a_k =+ \frac{T}{2}c_{kk}Y_k^2 = \frac{T}{2}c_{kk} \left[\tau^*_{k}\sum_{i=1}^{m} \sum_{j=1}^{m}\gamma_{ij,k}   Y_i Y_j \right]^2 \geq 0 ~. \label{kquartic}
\end{equation}
This expression (not to be confused with $\Delta F$ from Eq.~(\ref{quadF})) is the free energy change for the fast variable considered as an isolated system, so by the second law of thermodynamics this must be a positive definite quadratic form.  
The fast sub-system can exist in a condition of greater-than-equilibrium free energy for sustained periods of time, as long as at least one of the relevant slow sub-systems remains out of equilibrium.  This sustained state would be impossible in thermodynamic equilibrium.  Indeed, it is possible that $\Delta F_{k}$ could be large enough, that while in equilibrium, such states could be sampled only for extremely short periods of time during very rare, extreme, fluctuations i.e. events that are often considered to be thermodynamically inaccessible.  This quartic form in the slow variables $\{X_i\}$ provides a good measure for how far the fast subsystem can be pushed away from equilibrium under sustained conditions.  It is natural to think of  $\Delta F_{k}$  as an energy budget that can be used to overcome activation barriers, perhaps towards a desirable reaction product.

\vspace{0.2in}

\textit{Theorem 1:}

If $\dot{F} |_{\{Y_k=0\}}$  is the total free energy production rate with all fast states held at their equilibrium values, and $\dot{F} |_{\{J_k=0\}}$  is the total energy production rate with all fast states quasistationary then    
\begin{equation}
\dot{F} |_{\{J_k=0\}} -\dot{F} |_{\{Y_k=0\}} = \dot{F}_{extra} \leq 0~,   \label{Thm1}   
\end{equation}
where
\begin{equation}
\dot{F}_{extra}=-T\sum_{k=m+1}^n c_{kk}^{-1} \tau^*_k\left(\sum_{i=1}^{m} N_{ik}Y_i\right)^2 ~.    \label{sigmaex}
\end{equation}
This theorem is proved in a very similar manner to Theorem 2 of Ref.~\cite{Patitsas2014}. 
Given that the total free power must always be negative, the physical meaning for this result is that the magnitude of the total free power is made larger when the fast variables are allowed to relax by moving away from equilibrium values to quasistationary status.  Thus, the system as a whole approaches equilibrium sooner than without the relaxation.
Since I have shown that allowing the fast variables to relax to quasistationary states always leads to lower $\dot{F}$, I am led to formulate a variational principle which minimizes the free power in a certain sense. 

\section{Variational principles: discrete case} \label{sec:maxent}
 
I introduce the thermodynamic potential
\begin{equation}
\Phi \equiv  \sum_{i=1}^{m}{F}_i -\sum_{k=m+1}^n r_k^{-1} {F}_k  ~. \label{PhiDef}
\end{equation} 
The minus sign creates a clear distinction between slow and fast variables.  Fast variables are weighted in such a way that variables with small $r_k$ values receive more prominence.  For the dynamical problem it is the time derivative of $\Phi$ that is most important:
\begin{equation}
\dot{\Phi} \equiv  \sum_{i=1}^{m}\dot{F}_i -\sum_{k=m+1}^n r_k^{-1} \dot{F}_k = -T\sum_{i=1}^{m}J_i Y_i +~T\sum_{k=m+1}^n r_k^{-1}J_k Y_k = \dot{F}_{slow} +~T\sum_{k=m+1}^n r_k^{-1}J_k Y_k~. \label{PhiDotDef}
\end{equation} 
This function clearly differs, in general, from the total free power given in Eq.~(\ref{Fdot}), and it is used to formulate the following theorem.

\vspace{0.2in}

\textit{Theorem 2: (principle of minimal $\dot{\Phi}$)}

When $\dot{\Phi}$ is minimized, the fluxes for all the fast variables vanish, i.e., $J_k=0$ for $k>m$.

\vspace{0.1in}

To prove this theorem I first substitute into Eq.~(\ref{PhiDotDef}) for the fluxes, using Eq.~(\ref{Neqs}) to give 
\begin{equation}
\dot{\Phi} = -T\sum_{i=1}^{m}\sum_{j=1}^{m}L_{ij} Y_i Y_j +2T \sum_{i=1}^{m}\sum_{k=m+1}^n r_k^{-1} N_{ki} Y_i Y_k +T\sum_{k=m+1}^n r_k^{-1} L_{kk} Y_k^2   ~. \label{PhiDot2}
\end{equation}
Minimizing with respect to each fast variable, and recalling that the $N_{ik}$ coefficients do not depend on any fast variables $Y_k$, results in $n-m$ conditions:
\begin{eqnarray}
\frac{\partial \dot{\Phi}}{\partial Y_k} =  2T r_k^{-1} \sum_{i=1}^{m} N_{ik} Y_i +2T r_k^{-1} L_{kk} Y_k = 2T r_k^{-1} J_k =0~,  \label{diPhidY}
\end{eqnarray}
where comparison to Eq.~(\ref{Jk}) has been made.  Thus, I have proven that quasistationary states minimize $\dot{\Phi}$.  One more derivative with respect to $Y_k$ suffices to show that $\dot{\Phi}$ is minimized, since $L_{kk}>0$.  When all of the $n-m$ fast states are quasistationary then $\dot{\Phi}=\dot{F}_{slow}=\dot{F}$.  As shown in Section~\ref{statstates}, this total free power with all $J_k=0$ is less (but larger in magnitude) than it is with all $Y_k=0$.  Returning to the thermodynamic potential $\Phi$ and looking at the change  
\begin{equation}
\Delta\Phi\equiv \Phi_f -\Phi_i = \int_{t_i}^{t_f}\dot{\Phi}~dt~,
\end{equation}
one notes that if the Euler-Lagrange method is used and one integrates over time the general functional $\dot{\Phi}(\{Y_k\},\{\dot{Y}_k\},t)$, then the Euler-Lagrange equations \cite{Arfken} are 
\begin{equation}
\frac{\partial \dot{\Phi}}{\partial Y_k} =0~.
\end{equation}
Thus, by comparing to Eq.~(\ref{diPhidY}), the quasistationary states also minimize $\Delta\Phi$, i.e., by making it as negative as possible.  If, for example the initial value $\Phi_i$ is held fixed, then the final value $\Phi_f$ will be minimized. 

\vspace{0.2in}

\textit{Corollary 1: (principles of minimal $\dot{F}_{slow}$ and $\Delta F_{slow}$)}

One can easily show, using the method of Lagrange multipliers, that the quantities $\dot{F}_{slow}$ and $\Delta F_{slow}$ (from $t_i$ to $t_f$) are minimized when the fast variable functions $Y_k$ are quasistationary, if one also adds the $n-m$ constraints: $\dot{F}_{k}=0$ for $k>m$, i.e. for all fast states ($\dot{F}_k=0$ implies $\Delta F_k=0$).  The Lagrange multipliers are identified as $r_k^{-1}$.  

\vspace{0.2in}

\textit{Corollary 2: (principles of minimal total free power and $\Delta F$)}

Also, both the  total free energy production rate $\dot{F}$ and $\Delta F$ are minimized when the fast variables $Y_k$ take their quasistationary values, with the same $n-m$ constraints: $\dot{F}_{k}=0$ for $k>m$.  In this case, the Lagrange multipliers are $1+r_k^{-1}$.  The quasistationary states are very important since they minimize the total free power, as long as one understands the relevant constraints and that only fast variables are involved in the minimization procedure. In this sense, one may now refer to quasistationary states also as states of minimum free power.

Thus far I have formulated maximum entropy production principles in three ways.  A fourth formulation is obtained as follows. Taking the limiting procedure where slow state variables are actually fixed gives the following principle: when a system described by n variables is held in a state with fixed $Y_1,~Y_2,...,Y_m$ (with $m<n$) and minimum $\dot{\Phi}$, then the fluxes $J_k$ with $m<k\leq n$ vanish.   

When all of the $n-m$ fast states are stationary or quasistationary, the minimal value of the rate $\dot{\Phi}$ is given by Eq.~(\ref{PhiDotDef}):   
\begin{equation}
\dot{\Phi}_{min}=\dot{F}_{slow}|_{\{J_k=0\}}=\dot{F}|_{\{J_k=0\}} =-T\sum_{i=1}^m\sum_{j=1}^m L_{ij}Y_i Y_j+ \dot{F}_{extra}  ~.   \label{SigmaMax}   
\end{equation}
Physically, one thinks now of more than just fast variables relaxing to nonequilibrium values.  One thinks of the fast variables adjusting themselves so that the whole system gets to equilibrium faster.  In fact, while in the quasistationary states the whole system approaches equilibrium as fast as possible, given some reasonable restrictions. The extent of the adjustment of the fast variables must have limitations since minimizing $\dot{F}_{slow}$ or $\dot{F}$ without any constraints on the forces $Y_k$ would give an unphysical runaway result.  Instead, the fast variables relax until they become quasistationary and an important dynamical balance is achieved.  Mathematically, this balance is caused by the minus signs in front of the fast variable $\dot{F}_k$ terms in Eq.~(\ref{PhiDef}).  These minus signs emphasize the physical distinction between DR states and gate states.  The results presented here and in Ref.~\cite{Patitsas2014} cannot be arrived at without making this important distinction.  While the fast variables adjust themselves so that the whole system gets to equilibrium faster, they may spend considerable time in states with higher free energy than their equilibrium states ($a_k=0$) would have, i.e., $\Delta F_k>0$.  As discussed in Ref.~\cite{Patitsas2014}, one way this could happen in general is by heat flow, with the gate either getting cooler or hotter, in order to facilitate greater conductance of the DR subsystem.  Under isothermal conditions, particle transport into or out of the gate could be the way the DR conductance is increased.  This could provide for an effective method to trap particles.  Since this trapping is essentially entropic in origin it would have much in common with reported cases of entropic traps~\cite{Zwanzig1992a,Zwanzig1992b,PhysRevE.64.061106,PhysRevLett.96.130603,Asher1999,Craighead1999}. 

In magnetic systems, the magnetization of the gate could be the key variable controlling the DR conductance.  In general, the gate subsystem may exhibit more order (short-range or long-range) than in the equilibrium state which may be both uniform and lacking in order.  Thus, as alluded to in Sec.~\ref{statstates}, complex and otherwise difficult to attain states and interesting structures might be created as the fast variables sample their phase space and seek configurations that minimize the free power of the slow system (with the constraints $\dot{F}_{k}=0$ for $k>m$).  In particular, as a gate variable leaves equilibrium it may make a transition from a disordered state to one of order.  I term this interesting possibility an order-by-induction (OBI) transition.  

There are some common features when comparing OBI to the well-known phenomenon of order-by-disorder (OBD), an important concept in the field of frustrated magnetism~\cite{PhysRevLett.109.167201,Barnett2012,Villain1980,PhysRevB.88.220404}.  In a system with large accidental degeneracies that would otherwise have no long-range order, such order is formed if a subset of the degeneracies dominates because they support large fluctuations.  Fluctuations are important so that microstates are sampled properly.  Fluctuations of the gate variables are also important in TI so that the DRs can sample the different gate configurations properly, thus giving more statistical weight to higher conductance configurations. This importance is manifested by the presence of $\tau_k^*$ fluctuation time factors in the key equations such as Eqs.~(\ref{Nind}, \ref{sigmaex}).  The statistical aspect makes these systems a type of entropic trap, i.e., the system gets trapped into long-range order.   There are other significant commonalities between OBI and OBD: 1) order is formed when unexpected, 2) entropy is the most important thermodynamic variable in both effects, though in OBD one aims to maximize entropy while in OBI the rate of entropy production is maximized.  In both cases one takes isothermal conditions and analyzes the Helmholtz free energy.  The internal energy $U$ plays a secondary role in both cases.  Finally, I suggest that the driving force towards a transition to order may be stronger and more robust in the nonequilibrium realm.  It may be easier to clearly identify OBI than it is for OBD which has yet to be unambiguously identified experimentally.  For equilibrium systems one cools the system and hopes that impurities do not get in the way, whereas in nonequilibrium systems one can push the system harder towards possible ordering by applying larger conjugate forces on the DRs.


I also point out the similarities between OBI and the popular concept of order through fluctuations (OTF) originated by Prigogine and Nicolis~\cite{Nicolis1977}.  Clearly, fluctuations play a key role in both concepts.  In OTF, nonlinear dynamics guides chemical and biological systems towards states that are described by terms such as self-organization, emergence and complexity.  It remains to be seen if the extra terms in the dynamical equations introduced by TI can also play such a role and guide the dynamics of complicated systems with many variables into the same types of emergent and biologically complex states.  If so, then the results presented here will provide thermodynamic insights into these processes, for example by predicting an overall free energy budget available to surmount any obstacles along the way.  

I conclude this section with a few words on stationary states.  In the stationary (static) limit, all time derivatives of variables are constant, and quasistationary states become truly stationary.  For example, a large capacitor is replaced by a constant voltage source, and the reservoir becomes infinite.  Of course, not all variables are constant, just their time derivatives.  Indeed, the total system entropy increases at a constant rate.  Though the system is out of equilibrium, there are notable similarities with the mathematical structure of equilibrium thermodynamics.  One would set $d\dot{\Phi}=0$ to solve nonequilibrium problems in much the same way that one sets $dF=0$ to solve problems in equilibrium thermodynamics.  Indeed, thermodynamics should have more accurately been termed thermostatics, a point raised by previous workers in thermodynamics~\cite{degroot}.  In this spirit I propose the term nonequilibrium thermostatics.  This term would apply to any analysis of systems while they are in their stationary states. 

\section{Case of two variables} \label{sec:2var}

To help illustrate these concepts, I consider the simplest system possible that exhibits isothermal thermodynamic induction, the case where $n=2$ and $m=1$, i.e., one slow variable acting as the DR, and one fast variable which throttles or acts as a gate for the transport process in the DR.  This is an important case to consider since it likely suffices to cover many applications of isothermal thermodynamic induction.  This case illustrates the essential features of the DR variable interacting with a variable that has its dynamics coupled to the dynamical reservoir.  In the simplest case, these two variables would be completely uncoupled except that the kinetic coefficient for the DR variable $(i=1)$ happens to depend on $a_2$.  This means the two variables are uncoupled up to linear order i.e. $c_{12}=c_{21}=0$ and $L_{12}=L_{21}=0$.  This must be the case since if, for example, $c_{12}$ was nonzero, then one could not have one very slow timescale $\tau_1$ and one very fast timescale $\tau_2$.    The coupling at the nonlinear level will be described by the coefficient $\gamma_{11,2}$.  Note that $\gamma_{12,1}=\gamma_{21,1}=\gamma_{12,2}=\gamma_{21,2}=\gamma_{11,1}=\gamma_{22,1}=\gamma_{22,2}=0$.  Thus, under these assumptions, only $\gamma_{11,2}$ can be nonzero.   For the slow variable, Eq.~(\ref{relateIndDir}) and Eq.~(\ref{Nind}) gives $N_{12}=-\gamma_{11,2} g_{22}^{-1}Y_1$, which creates the coupling between the two variables, and Eq.~(\ref{Neqs}) becomes
\begin{equation}
{{\dot{a}}}_1= L_{11}Y_1  -  \gamma_{11,2} c_{22} Y_1  Y_2~,   \label{2vardyneq1}
\end{equation}
and
\begin{equation}
{{\dot{a}}}_2= L_{22}Y_2 + r_2\gamma_{11,2} c_{22} Y_1^2 ~. \label{2vardyneq2}
\end{equation}
Equation~(\ref{2vardyneq2}) is important for determining stationary states and can be rewritten, using Eq.~(\ref{gLtauOne}), as 
\begin{equation}
{{\dot{a}}}_2= L_{22} \left[ Y_2 + \tau_2^*\gamma Y_1^2 \right] ~. \label{2vardyneq2b}
\end{equation}
The first term in Eq.~(\ref{2vardyneq2}) describes a fast relaxation process while the second term describes the slowly varying induction effect, since $Y_1$ is slowly varying.

\subsubsection{Fast variable and slow variable}

It is instructive to connect my dynamical equations to van Kampen's general theory for separating fast variables from slow variables~\cite{VanKampen1985}.  Making use of Eqs.~(\ref{Xga}, \ref{gLtauOne}) allows Eqs.~(\ref{2vardyneq1}, \ref{2vardyneq2}) to be written as
\begin{equation}
{{\dot{a}}}_1= -\frac{a_1}{\tau_1}  -  \frac{\gamma}{c_{11}} a_1  a_2  ~, \label{2vardyneq1b}
\end{equation}
and
\begin{equation}
{{\dot{a}}}_2= -\frac{a_2}{\tau_2} + \frac{r_2\gamma c_{22}}{c_{11}^2} a_1^2 ~, \label{2vardyneq2b}
\end{equation}
where $\gamma_{11,2}\equiv\gamma$.  Noting $\epsilon\equiv \tau_2/\tau_1$ as van Kampen's small expansion parameter, and setting $y\equiv a_1$, $z\equiv a_2$ gives
\begin{equation}
{\dot{y}}= g(y,z) = -\frac{y}{\tau_1}  -  \frac{\gamma}{c_{11}} yz  ~, \label{geq}
\end{equation}
and
\begin{equation}
{\dot{z}}= \frac{1}{\epsilon} h(y,z) = -\frac{z}{\tau_2} + \frac{r_2\gamma c_{22}}{c_{11}^2} y^2 ~, \label{heq}
\end{equation}
where $g(y,z) \equiv -\frac{y}{\tau_1}-\frac{\gamma}{c_{11}} yz$, and $h(y,z) = -\frac{z}{\tau_1} + \frac{r_2\gamma c_{22}\tau_2}{c_{11}^2\tau_1} y^2$.
The gate variable is expanded as
\begin{equation}
z=z^{(0)}+\epsilon z^{(1)} + \epsilon^2 z^{(2)} +...~.
\end{equation}
Setting $h(y,z^{(0)})=0$ is the condition for a quasistationary state. From Eq.~(\ref{heq}) I determine that
\begin{equation}
z^{(0)} = \frac{r_2\gamma c_{22}\tau_2}{c_{11}^2} y^2 \equiv \phi(y)~.  \label{z0}  
\end{equation}
Explicitly this is a function of $y$ only.  By inserting $z^{(0)}$ into Eq.~(\ref{geq}) I obtain
\begin{equation}
{\dot{y}}^{(0)}= g(y,z^{(0)}) = -\frac{y}{\tau_1}  -   \frac{r_2\gamma^2 c_{22}\tau_2}{c_{11}^3} y^3~,
\end{equation}
which finishes the dynamical problem to zeroth order.  To the next order in $\epsilon$
\begin{equation}
{\dot{z}}^{(1)}= h_z(y,z^{(0)}) z^{(1)}~.   \label{z1}
\end{equation}
In general $h_z$ must be negative to ensure stability~\cite{VanKampen1985} and indeed, in my case, $h_z=-1/\tau_1$ is always negative.  I use Eq.~(\ref{z1}) to isolate
\begin{eqnarray}
z^{(1)}&=&\frac{1}{h_z}\frac{dz^{0}}{dt}=\frac{1}{h_z}\frac{d\phi}{dy}\frac{dy}{dt}=\frac{\phi_y g(y,\phi(y))}{h_z}\equiv\psi(y) \\
       &=&  \frac{2r_2\gamma c_{22}\tau_2}{c_{11}^2\tau_1^2} y^2   \left( 1 + \frac{r_2\gamma^2 c_{22}\tau_1\tau_2}{c_{11}^3} y^2 \right)~,
\end{eqnarray}
thus demonstrating the entrainment of the gate variable while in the quasistationary state, i.e., $z=a_2$ is specified in terms of the slowly varying $y$ variable.  To first order the dynamics for $y$ is given by
\begin{equation}
{\dot{y}}^{(1)}= g(y,\phi(y)) +\epsilon\psi(y)~,
\end{equation}
which explicitly shows that the fast variable has been eliminated exactly as prescribed by van Kampen's procedure~\cite{VanKampen1985}.  Though van Kampen did not predict TI, I have verified that the terms introduced by TI are accommodated by his general approach.

\subsection{Fast (gate) variable}

\begin{figure}[ht]
	\centering
		\includegraphics[width=0.4\textwidth]{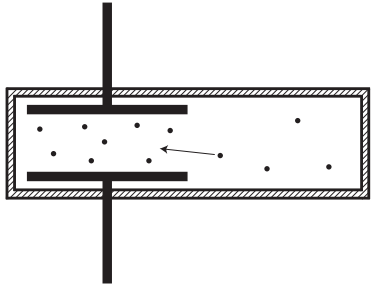}
\caption{\label{fig:2boxes} Schematic diagram of a fluid-filled chamber depicting an induced transverse electromigration effect.  Electrical potential is applied across the conducting plates shown on the left side of the chamber.  Dissolved ions are induced (see arrow) to move into the space between the plates.}
	\label{fig:plates}
\end{figure}

In this work, both subsystems exhibit processes that involve the exchange of particles.  I discuss subsystem 2 first.  The transport process for the subsystem 2 will be particle diffusion.  One may imagine a cell or chamber, such as depicted in Fig.~\ref{fig:plates} where particle transfer is between the left and right halves of the chamber.  Specifically, $x_2$ could be the number $N_2$ of particles on the left half, and $a_2=\Delta N_2=N_2-\bar{N}_2$ is the change from the equilibrium value which denote as $\bar{N}_2$.  Using Eqs.~(\ref{quadF}), (\ref{Yp}) and (\ref{Xga}), the conjugate force can be related to the difference in chemical potential between the two halves as
\begin{equation}
Y_2=-\frac{1}{T}\Delta \left(\frac{\partial F}{\partial N_{2}}\right) = -\frac{1}{T}\Delta\mu_2~.  \label{Y2}
\end{equation}
Also note that

\begin{equation}
c_{22}^{-1}=\frac{1}{T}\left(\frac{\partial\mu_2}{\partial N_2}\right)_T~.   \label{g22}
\end{equation} 
For convenience I define an intensive and dimensionless parameter $\zeta$ as
\begin{equation}
\zeta\equiv\frac{\bar{N_2}}{k_B T}\left(\frac{\partial\mu}{\partial N_2}\right)_T = \frac{\bar{N_2}}{k_B } c_{22}^{-1}~. \label{zeta}
\end{equation}
Note that $\zeta=1$ for a classical ideal gas as well as for a dilute solute.  In order to ensure (equilibrium) thermodynamic stability this parameter cannot be negative.
If the characteristic length scale of subsystem 2 is $w$ then $\tau_2=w^2/D$ where $D$ is the diffusion coefficient.  Making use of Eq.~(\ref{gLtauOne}) and Eq.~(\ref{zeta}) the Onsager coefficient $L_{22}$ is related to $D$ as
\begin{equation}
L_{22}=\frac{c_{22}}{\tau_2}=\frac{D \bar{N}_2}{\zeta k_B w^2}~.
\end{equation}
Using Eq.~(\ref{2vardyneq2}), the Helmholtz free power for the gate variable is
\begin{equation}
\dot{F}_{G}=\dot{F}_2= -T{\dot{a}}_2 Y_2 = -T L_{22}Y_2^2  -T  r_2\gamma c_{22} Y_1^2  Y_2~. \label{Fdotfast3}
\end{equation}

\subsection{Slow (DR) variable}

The transport process for the slow variable 1 is specifically electrical conduction by transfer of charged carriers.  In particular I take subsystem 1 to be the anode of an electrical circuit with large $RC$ time constant.  For example, the anode could be the top plate shown on the left side of Fig.~\ref{fig:plates}.  I take $N_1$ to be the number of electrons on the anode, so that the electrical charge $Q=-e\Delta N_1$ where $e=+1.602\times 10^{-19}$ C.   If the applied electric potential difference is $W$ then $\Delta\mu_1=-eW$~, and by Eq.~(\ref{Yp})
\begin{equation}
Y_1=+\frac{e}{T}W~.  \label{Y1toW}
\end{equation}
Note that if $W$ is positive then the anode is at positive potential and one would expect the anode to receive electrons and $\dot{a}_1$ to be positive.  This is consistent with Eq.~(\ref{2vardyneq1}).
  The electrical charge $Q=-e\Delta N_1$, and by Eq.~(\ref{Xga}) 
\begin{equation}
W=\frac{T}{e^2 c_{11}} Q~,
\end{equation}
 and one concludes that $\frac{T}{e^2}c_{11}^{-1}=1/C$. 
Large capacitance will make $c_{11}$ large and subsystem 1 slow.  The electrical current $I=e\dot{a}_1$ so if Eq.~(\ref{2vardyneq1}) is re-expressed as $\dot{a}_1=M_{11}Y_1$ and then comparison is made to Ohm's law then
\begin{equation}
M_{11}=\frac{T}{e^2}G~,    \label{MtoG}
\end{equation}
 where $G=1/R$ is the electrical conductance of subsystem 1.  For the induction coefficient defined by Eq.~(\ref{gammaijk}) the result is
\begin{equation}
\gamma=\gamma_{11,2}=\frac{T}{e^2}\frac{\partial G}{\partial N_2}~. \label{gamm112}
\end{equation}
The timescale for subsystem 1 is $\tau_1=RC$ and one readily verifies that Eq.~(\ref{gLtauOne}) holds.  The electrical power $P_1\equiv P_R$ dissipated into the DR circuit is 
\begin{equation}
P_R= IW= TY_1\dot{a}_1=-\dot{F}_1~.
\end{equation}
Using Eq.~(\ref{2vardyneq1}), the Helmholtz free power for the DR is
\begin{equation}
\dot{F}_{R}=\dot{F}_1= -T{{\dot{a}}}_1 Y_1 = -T L_{11}Y_1^2  +T  \gamma c_{22} Y_1^2  Y_2~.  \label{Fdotslow3}
\end{equation}
It is helpful to define
\begin{equation}
\dot{F}_0\equiv  -T L_{11}Y_1^2 =-G\Phi^2~,
\end{equation}
which is the rate of change of the total free energy when $Y_2=0$, i.e., $\dot{F}_0=\dot{F} |_{\{Y_2=0\}}$.  Thus $\dot{F}_0=-P_0$ where $P_0$ is the power $P_1$ dissipated in the electrical circuit, when $Y_2=0$.


\subsection{Quasistationary state}

The induction term in Eq.~(\ref{2vardyneq1}) manifests itself as $+r_2\gamma_{11,2} c_{22} Y_1^2$ and affects the dynamics of $a_2$.  Thermodynamic equilibrium corresponds to $a_1=a_2=0$.  If the DR variable $a_1$ is held away from zero, one sees that $a_2$ will be induced to also move away from zero.  Afterwards the diffusion term contributes and opposes induction.  Over a timescale of $\tau_2$ a balance between induction and diffusion is achieved and the system becomes quasistationary.  
The single quasistationary state available for this system comes from setting $\dot{a}_2=0$.  Thus the function $\dot{\Phi}(Y_1,Y_2)$ is maximized by holding $Y_1$ constant while varying $Y_2$.  From Eq.~(\ref{2vardyneq2b}) one obtains:
\begin{equation}
 Y_{2,qss} =- \tau_2^*\gamma Y_1^2  ~, \label{Var2SS}
\end{equation}
or, using Eq.~(\ref{Y2})
\begin{equation}
 \Delta\mu_{2,qss} = \frac{\tau_2^*\gamma}{L_{11}} P_0  ~. \label{Var2SSc}
\end{equation}
Equation~(\ref{Var2SSc}) is helpful towards understanding how TI might affect the example systems discussed below in Section~\ref{sec:examples}.  Before proceeding to these examples some further discussion is warranted.  For convenience I define another intrinsic and dimensionless variable proportional to $\gamma$ as
\begin{equation}
\kappa\equiv\frac{\bar{N}_2 \gamma}{L_{11}}~.\label{kappa1}
\end{equation}
Using Eq.~(\ref{kappa1}) and using the subscripts 2 and G interchangeably:
\begin{equation}
 \Delta\mu_{G,qss} = \kappa \frac{P_0\tau_2^*}{\bar{N}_2} = \kappa \frac{U^*}{\bar{N}_2}  ~, \label{Var2SSc2}
\end{equation}
where $U^*\equiv P_0\tau_G^*$ is the energy dissipated by the transport of variable 1 during the time step $\tau_G^*$.
This simple relation is quite useful in analyzing potential applications of thermodynamic induction when $U^*$ is known.  Another helpful result arises from defining an effective temperature as
\begin{equation}
T_{eff}\equiv \frac{\Delta\mu_{G,qss}}{k_B} = \frac{\kappa U^*}{\bar{N}_2 k_B}   ~. \label{Teff}
\end{equation}
This equation is useful in ascertaining the temperature at which thermodynamic induction would be expected to be observed.  For example, if the actual temperature $T$ is much larger than $T_{eff}$ then induction effects would be very difficult to observe. 
The induction effect will result in subsystem 2 particles aggregating between the electrodes of subsystem 1 if $\gamma$ is positive.  I use Eq.~(\ref{agX}) to obtain an expression for $\Delta N_2$, while combining with Eq.~(\ref{Var2SSc2}) and making use of Eq.~(\ref{zeta}).  The result is  
\begin{equation}
\Delta N_{G,qss}=\frac{\bar{N_2}}{k_B T}\zeta^{-1}\Delta\mu_{G,qss} = \zeta^{-1}\kappa \frac{U^*}{k_B T}   ~. \label{DeltaN2qss}
\end{equation}

From Eqs.~(\ref{Nind}),~(\ref{sigmaex}), and (\ref{Var2SSc}) one identifies the quasistationary quantity which plays an important role in Theorem 1:
\begin{equation}
\dot{F}_{extra} = -\frac{\kappa^2 U^*}{\zeta\bar{N}_2 k_B T}P_0 ~. \label{dotFextra2var} 
\end{equation}
By Eq.~(\ref{Thm1}), Theorem 1 can be expressed here as $\dot{F}_{\{J_G=0\}}-\dot{F}_{\{Y_G=0\}}=\dot{F}_{extra}<0$.  The same result holds for the slow variable, i.e., for $\dot{F}_{R}$, since $\dot{F}_{G}=0$ for the quasistationary state.  The simple expression in Eq.~(\ref{dotFextra2var}) helps to illustrate the important result, i.e., the single DR variable will approach equilibrium faster because of thermodynamic induction, regardless of the sign of $\gamma$.  Specifically the relative increase in this rate of energy transfer is $\frac{\kappa^2 U^*}{\zeta\bar{N}_2 k_B T}$.  This is seen by inspecting Eq.~(\ref{dotFextra2var}) and noting that $-P_0$ is the rate when $\gamma=0$.

With careful inspection of Eq.~(\ref{Fdotfast3}) one notes that while the gate variable is quasistationary, there is a balance between two terms. The first term is purely dissipative and is given by $r_2 \dot{F}_{extra}$ and tends to push the gate subsystem towards lower $F_G$.  Put another way, this term pushes the gate towards sampling all of its microstates equally (equilibrium). The second term is more interesting as it is 
an inductive anti-dissipative process that has the tendency to push the free energy of the gate subsystem upwards, i.e., 
\begin{equation}
\dot{F}_{G,ind,qss}=-r_2\dot{F}_{extra}>0~.  \label{selectionPower}
\end{equation}
I am not aware of any process with this property in the realm of equilibrium thermodynamics.  It is interesting to speculate as to the physical interpretation of the power term $\dot{F}_{G,ind,qss}$.  It seems to push the gate towards sampling microstates in a selective manner, i.e., by giving preferential weight towards those microstates that result in higher DR conductance.

If one asks how much gate variable free energy could be garnered from TI, the answer is supplied by Eqs.~(\ref{kquartic}) and (\ref{Var2SSc2}) as
\begin{equation}
\Delta F_{G}=-\frac{1}{2}\dot{F}_{extra}\tau_2^*= \frac{\kappa^2 (U^*)^2}{2\zeta \bar{N}_2 k_B T} ~.  \label{kquartic2}
\end{equation}
Note that apart from the fluctuation timescale $\tau_2^*$, this energy differential does not depend on the details of the gate.  One may think of this energy then as a budget allocated towards the gate, as a means towards getting the DR to equilibrate more quickly.  This energy could, for example, be used to directly offset any activation energy $U_A$ for the gate as a whole, i.e., not per particle.  The energy offset could be implemented in the relevant Boltzmann factor, i.e., where $\exp{[-U_A/k_B T]}$ gets replaced, and enhanced, by $\exp{[-(U_A-\Delta F_G)/k_B T]}$.

I conclude this section with further discussion of the physical significance of the free energy $\Phi$ and the minus signs in Eqs.~(\ref{PhiDef}) and~(\ref{PhiDotDef}).  An attempt to deposit an amount of energy $E$ into the dynamical reservoir only would not succeed.  Thermodynamic induction would cause a fraction of this energy to be shunted into the gate (or in general any coupled fast variables).  This is very similar to the physical interpretation of thermodynamic free energy potentials such as the enthalpy, Helmholtz free energy and Gibbs free energy. For example, with isobaric systems governed by the enthalpy function, some fraction of any energy deposited into a system is used up in accommodating changes in the system volume.  The situation here is similar, but I emphasize that in the work presented here, the systems are out of equilibrium.

\section{Example Systems}\label{sec:examples}

Below, I consider four examples all with one slow and one fast variable.  The analysis leans heavily on the basic result for the induced shift in chemical potential, as expressed in Eqns.~(\ref{Var2SSc},~\ref{Var2SSc2}).  I believe that all four systems can lead to experiments that could verify or refute my claims for the existence of thermodynamic induction.  These examples also emphasize the broad applicability of TI to several systems important in physics and chemistry, and I anticipate that TI would also be of interest to researchers working on analytical purification of solvents, superfluids and their applications, semiconductor devices, semiconductor crystal growth, as well as to surface scientists.

\subsection{Solute conduction and induced electromigration}

If two electrodes are immersed in an ionic solution and biased, then thermodynamic induction would drive ions, both positive and negative, towards the region between the electrodes.  This effect is illustrated in Fig.~\ref{fig:plates} with the plates occupying the left half of the electrochemical cell.  Applications related to the purification of fluids are evident.

The coefficient $\gamma$ is positive for both positive and negative ions since in both cases the conductance is enhanced when ions move into the space between anode and cathode.  If the conductance varies linearly with ionic concentration then $\kappa=+1$.  In this linear case the power $P_R$ dissipated in the electrical circuit will also scale with $N_G$.  By Eq.~(\ref{Var2SSc2}), while in the quasistationary state, the change in chemical potential for ions $\Delta\mu_{G,qss}$ will not depend on $N_g$, so there is equal effectiveness at high and low ionic concentrations.
It is important to point out the strength of the induction by calculations with specific numbers.  An important parameter that points towards a very small effect is the timescale $\tau_G^*$ which will take a value near $10^{-13}$ s~\cite{Zangwill}.  If one considers a standard-sized electrochemical cell, a cube 2.5 cm on each side, holding a 0.10 M NaCl solution, with 10 W of electrical power running through it, then Eq.~(\ref{Var2SSc2}) predicts $\Delta\mu_{G,qss}=5.3\times 10^{-34}$ J.  This is a very small number and would be very difficult to observe since it is about 13 orders of magnitude smaller than $k_B T$ at room temperature.  The reasons are the small value of $\tau_G^*$ as well as the large value of $N_2=1.9\times 10^{21}$ ions for this example.  Thermodynamically induced electromigration will not be easy to observe in standard electrochemical cells under normal conditions.  

Under more extreme conditions, heat dissipation will be a problem if one tries to increase the electrical current, so it may be more wise to look at systems with lower carrier densities and higher electric fields.
Expressing $P_0$ in Eq.~(\ref{Var2SSc2}) as $P_0=G W^2$, then in terms of the electric field, $E$:
\begin{equation}
 \Delta\mu_{G,qss} =\kappa \frac{G W^2\tau_2^*}{\bar{N}_G} = \kappa e u  E^2\tau_2^*   ~, \label{Var2SSc3}
\end{equation}
where the conductance is expressed in terms of the carrier mobility $u$.   
For the case of water at room temperature, H$^+$ ions have a relatively high mobility value of $3.7\times 10^{-7}$ m$^2/$Vs~\cite{noauthororeditor2007handbook}.  Also, the maximum possible electric field before dielectric breakdown is 70 MV/m.  Assuming this maximum electric field, $\Delta\mu_{G,qss}=1.8\times 10^{-4}$ eV, or about 140 times smaller than $k_B T$.  By Eq.~(\ref{DeltaN2qss}) this would give a $\Delta N_G/N_G$ ratio of 0.7~\% which should be within the detection threshold of standard analytical techniques.  For example, if the volume of the solvent outside of the cell is the same as that of the cell, then the electrical conductivity of the solvent outside the cell should be reduced by $0.7~\%$.    If the same mobility could be achieved in benzene at maximum electric field, then $\Delta\mu_{G,qss}$ gets up to just $25$ times smaller than $k_B T$.  This would give a $\Delta N_G/N_G$ value of 4~\% which would be quite easy to measure, provided there is effective subtraction of competing effects such as Coulomb attraction to the plates and dielectric attraction to strong fields.     
These tests could be conducted in fluid cells depicted in Fig.~\ref{fig:plates} with a lateral dimension of 0.02 m.  A second set of plates could be added to the right side of Fig.~\ref{fig:plates}.  These would be used for measuring conductance under small signal, very low power conditions (TI insignificant).  One would observe a decrease in electrical conductivity on the right side when high voltage is applied across the left plates.  
A more exotic system to explore is superfluid helium where the mobility of dissolved ions can be very high, for example values approaching 1 m$^2/$Vs for ions in liquid He II at 0.5 K~\cite{Reif1961}.   In this system, electromigration by TI (as illustrated in Fig.~\ref{fig:plates}) should be observable at electric fields much smaller than the dielectric strength.  

\subsection{Carrier conduction in semiconductors}

Mobilities of charge carriers in doped semiconductors can also be very high.  For example for conduction band electrons in Si at room temperature, $u=0.14$ m$^2/$Vs.  Also, $\tau_G^*\approx~1$~ps, a quantity determined by electron-phonon scattering rates~\cite{HarrisonSolid, AM}.  By use of Eq.~(\ref{Var2SSc3}) I find that the level of $\Delta \mu_{G,qss}=k_B T$ will be achieved at an electric field of $4.3\times~10^{5}~$V/m, which is about 100 times smaller than the dielectric strength of Si. Put another way, by Eq.~(\ref{Teff}), at these electric fields the effective temperature is equal to room temperature.  Thus, significant induction effects should be detectable in high field regions of semiconductor systems fabricated to resemble what is depicted in Fig.~\ref{fig:plates}.  These systems do not have to be small and could be built at the scale of 0.01 m or so.  My results should also apply to smaller length scales which would include the length scales of technologically important semiconductor devices.  For example, near the source and drain electrodes of a MOSFET, I predict charge carriers to be induced in significant numbers to small areas of highest electric field.

\subsection{Dopant diffusion}

Equation~(\ref{Var2SSc3}) also applies to dopant atoms embedded into the crystal matrix of semiconductors.  As dopants slowly diffuse towards a region with higher electric field the carrier concentration (associated with the dopant ionization i.e. donating or accepting) and hence the electrical conductivity of the region will increase.  Thus $\kappa=1$, $\tau_G^*=0.1~$ ps, and the mobility $u$ in Eq.~(\ref{Var2SSc3}) will be that of the carrier.  For example, for P diffusion in Si, the donated carrier mobility at room temperature will be $0.14$ m$^2/$Vs.  Thus, at electric fields of $4.3\times~10^{5}~$V/m one would expect $\Delta \mu_{G,qss}$ to be near 10~\% of $k_B T$ and therefore have an appreciable effect on the doping profile.  The biggest drawback is that it could take a very long time to achieve the quasistationary state.  This diffusive timescale is $\tau_G\approx b^2/D$ where b is the crystal lattice constant and $D$ is the diffusion coefficient for the dopant.  Because of the large energy barrier for thermally activated atomic diffusion, $D$ typically has very small values.  Even at 1100~K the value is approximately $10^{-20}$~m$^2/$s for P-Si~\cite{Sze}.

If one is able to wait long enough, perhaps for a period of several years, TI should produce significant changes to the dopant density profile.  These profiles could be directly measured both before and after using analytical techniques such as secondary ion mass spectrometry~\cite{RCM:RCM4046}. 
Also,	as the dopants diffuse one would expect to see the conductance of the device increase over time.  There could be a tendency for a runaway effect: as the conductance increases the temperature rises which in turn increases the dopant diffusion rate.  This suggests that there may very well be a link between thermodynamic induction and semiconductor device failure.

\subsection{Surface diffusion and scanning tunneling microscopy}

It is well known that surface adsorbates can strongly affect the conductance of the tunnel junction of an STM.  If the adsorbates are freely diffusing across the surface then one might expect adsorbates to be attracted to the region underneath the tip of an STM (or repulsed if the adsorbate reduces the junction conductance).  In order to analyze this system I incorporate the tunnel junction conductance $G$ into Eq.~(\ref{Var2SSc}).  Using a logarithmic derivative gives
\begin{equation}
 \Delta\mu_{G,qss} = P_0\tau_G^*   \frac{d\ln G}{d N_G}  ~. \label{DeltamuSTM}
\end{equation}
The conductance of an STM tunnel junction can be treated using a square tunneling potential barrier~\cite{STMbook} which gives
\begin{equation}
G=G_0 e^{\alpha(-r+N_G a)} ~, \label{GSTM}
\end{equation}
where $a$ is the apparent height of the adsorbate as imaged by STM.  The apparent height can be a positive or negative quantity and plays an important role here.  Inserting Eq.~(\ref{GSTM}) into Eq.~(\ref{DeltamuSTM}) gives
\begin{equation}
\Delta\mu_{G,qss} = P_0\tau_G^* \alpha a = U^* \alpha a~. \label{DeltamuSTM2}
\end{equation}
For common tip materials such as W and Pt, $\alpha\approx 22$~nm$^{-1}$~\cite{STMbook}.  For surface diffusion I take $\tau_G^*=0.1$~ps~\cite{Eigler91,Zangwill}.
The reciprocal of the time constant $\tau_G^*$ can be the attempt frequency $\nu_0$ used in theoretical treatments of thermally activated reactions.  With an energy barrier $E_b$, such treatments implement this frequency in the rate expression 
\begin{equation}
R=\nu_0 e^{-E_b/k_B T} ~, \label{rate}
\end{equation}
and for the case of thermal diffusion then $\tau_G^*$ would equal $1/\nu_0$~\cite{Zangwill}.  Often $\nu_0$ corresponds to vibrational frequencies~\cite{Eigler91,Zangwill,Avouris93} though it should be noted that $\nu_0$ can vary widely from system to system~\cite{EstrupTully1986}.  For the systems to be discussed below, the inverse of the most relevant vibrational frequency (such as the adsorbate--substrate stretch mode) is often considerably larger than 0.1 ps.  Though it is tempting to use such larger times in Eq.~(\ref{DeltamuSTM2}) to bolster the argument for TI, I point out that this may be too optimistic.  Other modes and their frequencies must also play an important role in creating the thermal fluctuations that are the ultimate source of the TI effect.  Many of these timescales are likely at the 0.1 ps range, for example, the vibrational modes in the substrate below the adsorbate.  I choose then to be cautious and use the 0.1 ps value for $\tau_G^*$ throughout the discussion of surface adsorbates and their manipulation.

Note that the adjustable parameter in Eq.~(\ref{DeltamuSTM2}) which is at the disposal of the STM operator is $P_0$, i.e., the product of the applied bias and the tunnel current, both of which are readily controlled.  Usually the applied bias is directly set and the current is held near a setpoint value by using electronic feedback.  
Typical imaging conditions for STM on metal surfaces are 0.1 V applied bias, and 100 pA tunnel current i.e. $G=10^{-9}$~$\Omega^{-1}$. 
With these numbers, and assuming a value $a=0.1~$nm, I obtain the rather small value of $\Delta\mu_{G,qss}=~$0.014 meV. 
Under these benign conditions one would not expect TI to have a strong effect on surface atoms.  This is consistent with traditional understanding of STM, i.e., that under these conditions a tip can scan across the surface for several hours, while taking many images, without any observable disruption of the surface.  In order to create modifications to the surface and/or the tip one either increases the applied bias or the tunnel current, or both.  In any case, $P_0$ is increased.  If I increase the $P_0$ value in the calculation by a factor of 2000, then $\Delta\mu_{G,qss} = 28~$meV.  This number is definitely significant, being quite close to $k_B T$ at room temperature.  It is also in the same range as binding energies and diffusion barriers for physisorbed species.  For example if $\gamma>0$ then a potential well gets induced in the vicinity of the tunnel junction and the depth of this well is 28 meV.  Imaging adsorbates readily diffusing on a surface is challenging for STM as one does not view individual adsorbates but rather one observes a noisy ``cloud'' caused by the rapid motion and the relatively slow STM electronics.  This is the case for STM studies of benzene physisorbed at 95 K on Si(111)7x7~\cite{Wolkow1998}.  Even so, the presence of the adsorbate under the tip is readily detected by the STM circuitry as a current spike, and statistics could be obtained over a long time to ascertain if there is a significant change in the occupancy under the tip, relative to the benign conditions.  

If I push the STM junction parameters to extreme values, i.e., $\approx~$3 V applied bias and $\approx$ 300 nA current, and if the apparent height is $0.2~$nm, then $\Delta\mu_{G,qss} \approx 2.8~$eV.  Large apparent heights near 0.2 nm are not uncommon at high bias since bonding resonances become accessible when the tip Fermi level is pushed a few eV away from the sample Fermi level.  This is the case, for example, in STM studies of Cl chemisorbed onto Si(111)7x7~\cite{Boland1989}.  Near 3 V bias, the Cl atoms dissociate at a high rate and are imaged as very bright features, approaching 0.2 nm in apparent height~\cite{Patitsas2009}.  The $\Delta\mu_{G,qss}$ energy of 2.8 eV is very significant and is large enough to break almost all chemical bonds.  I would expect disruption of both the tip and even substrate atoms may be dislodged, and indeed this is what is observed; the Si(111)7x7 surface is disrupted at 3 V bias by the intense electric fields~\cite{Avouris1991}.  Also, the very high current level of 300 nA is difficult to achieve reliably with STM of Xe on Ni(110)~\cite{Avouris93}.  In fact, it is a well-known technique in STM operation to improve a poorly performing tip by subjecting it for a short time to either high currents, such as several hundred nA, or to high voltages near 5 V, or both.  Likewise one refrains from exposing a tip in good operating condition to these extreme settings.

These simple numerical calculations motivate the following proposal for future experiments seeking to directly verify TI.  One could establish an STM tunnel junction on a surface with quickly diffusing adsorbates.  One could then look for an increase in junction conductance when parameters are set for a relatively large $\Delta\mu_{G,qss}$ (say 100 meV) as compared to the conductance under (benign) conditions (say $\Delta\mu_{G,qss}=0.01~$meV).  One can also take this a step further: by collecting statistics on current spikes, as described above, one can conclude more than just mean values but also information on the rms noise and correlation functions.  To my knowledge such an experiment, though achievable with currently available variable temperature STM systems, has not yet been performed.  I believe that such an experiment giving a positive result would provide strong evidence for TI. 

\subsubsection{Influence of TI on manipulation of atoms and molecules by STM}

I briefly move away from my discussion of surface diffusion and instead analyze and review reported experiments involving manipulation of individual atoms and molecules by STM, and I hope to see if any of these may be interpreted in terms of thermodynamic induction.  I note, however, that these STM experiments were conducted at lower temperatures with adsorbates that are essentially frozen into surface binding sites.
	
\begin{figure}[ht]
	\centering
		\includegraphics[width=0.25\textwidth]{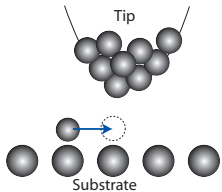}
	\caption{Schematic diagram illustrating a lateral diffusion event aided by TI.  If the diffusing species increases the tunnel junction conductance then I predict that the species will be induced to move underneath the tip.}
	\label{fig:Tip}
\end{figure}  

Xenon atoms physisorbed on cold metal surfaces have been studied extensively by various surface techniques, including STM.  At 4 K the atoms rarely move and at 0.1 V bias, they consistently appear as 0.16 nm high protrusions on the Ni(110) surface i.e. an apparent height $a=+0.16$~nm.  Zeppenfeld \emph{et al.} have demonstrated that on Ni(110) at 4K, STM can be used to pull Xe atoms across the surface with great effectiveness along $<1\bar{1}0>$ rows~\cite{Eigler92}.  This is done with the STM tip placed closer to the surface than is usual for routine imaging: in this case the threshold junction impedance is 4.8 M$\Omega$ i.e. $G=2.1\times~10^{-7}$~$\Omega^{-1}$.  During these remarkable manipulations, the Xe atom is essentially trapped in some sort of potential well centered underneath the tip, while still physisorbed to the Ni surface.  A satisfactory explanation for the trapping has not been established.  Van der Waals-type interactions between the Xe and the tip were suggested as the best explanation but this cannot be the case as such forces would certainly lead to the Xe jumping to the tip, and this is not observed.  At even smaller reported junction impedances, near 900 k$\Omega$, such perpendicular processes (jumps to tip) are indeed initiated~\cite{Eigler92} and these jumps have been attributed to multiple (ladder-type) vibrational excitations caused by the high tunnel current density~\cite{Avouris93}.  For the sliding process at 4.8 M$\Omega$ junction impedance, vibrational excitations would not result in an attractive potential well.  Currently accepted mechanisms for stimulating and breaking bonds, such as vibrational excitation, temporary ion resonances, etc., are not sufficient for trapping and controlling.  For example, chemisorbed benzene can be dislodged from a Si(100) surface using one of these mechanisms, but the dislodged molecules do not stay underneath the tip.  Rather, they are found elsewhere on the surface in subsequent scans, or not found at all~\cite{Patitsas2000}.  The sliding manipulation must involve some type of bond-excitation process as well as an attractive potential well. I believe that TI can supply the potential well part of this type of manipulation.
Using the Xe/Ni(110) parameters in Eq.~(\ref{DeltamuSTM2}) with $\tau_G^*=~$0.1 ps, I obtain $\Delta\mu_{G,qss}=4.7$ meV.  Thus an induced potential well should be created under the tip with a depth over ten times $k_B T$. As the tip is scanned laterally across the surface, the trapped Xe would follow along, thus explaining the Xe manipulation results nicely.   

This seems very promising as evidence for TI, but one must be careful in interpreting the results of Eq.~(\ref{DeltamuSTM2}).  In Eq.~(\ref{DeltamuSTM2}), it is assumed that the gate subsystem has had time to relax into the quasistationary state.  This requires a time of several $\tau_G$.  Given that a diffusion barrier for Xe/Ni(111) of 14 meV has been measured using optical diffraction~\cite{Nabighian2000}, at 4~K it seems unlikely to see the required relaxation on Ni(110) even after several hours have passed.  From the energetic point of view, however, the Xe/Ni(110) system is not far away from a fast enough relaxation time.  For surface diffusion, the diffusive relaxation time $\tau_{G}$ is often given by $1/R$, where $R$ is the rate in Eq.~(\ref{rate})~\cite{Zangwill}.  I have calculated that if the temperature were raised to 6 K, then $\tau_G$ would be lowered to 1s, and in this case the induced potential well would be easily observed in STM.  This modest temperature rise may be caused by the local heating (which could be highly local vibrational excitation~\cite{Avouris93}) caused by the tunnel current.  It is also possible that the diffusion barriers immediately adjacent to the tunnel junction are substantially lowered by the strong electric field due to the presence of the tip. 
For example, if the barrier is lowered to $E_b=~$10 meV then at 4~K, $\tau_G=1$~s, and again, the quasistationary state would readily be achieved.  Finally, one more plausible scenario is that the thermodynamically induced potential well profile, leads to lower diffusion barriers immediately adjacent to the tunnel junction.  Of course, the potential well needs to be present in the first case, so this explanation involves bootstrapping.  At this moment I merely point out that this effect would drive the mean value of $\tau_G$ to a lower value than otherwise.  In this sense then, the tunnel junction working in concert with TI, serves to catalyze the diffusion process.

This lateral type of manipulation by STM is not unique.  At 4 K, the same type of tip-assisted diffusion has been reported for physisorbed CO on Pt(111) at a threshold junction impedance of 300 k$\Omega$~\cite{Eigler92}.  In this case the effective height in STM is reported as $a=+0.5~$nm.  With these numbers I calculate  $\Delta\mu_{G,qss}=30$ meV, a deeper well.  
For both Xe and CO adsorbates, I propose that similar studies be conducted at higher temperatures near 50~K, where surface diffusion events for Xe or CO will be common, and $\tau_G$ will be small enough to assure that the quasistationary state is reached.  Under these conditions, under-the-tip occupation statistics, detected as current pulses, can be readily tallied.  

Further examples of demonstrated atomic and molecular manipulations by STM include results for Cu and Pb on Cu(211)~\cite{Meyer97} as well as for iodine and phenyl groups on Cu(111)~\cite{Meyer2000}.  For the study of Pb/Cu(211) conducted at 30 K, the individual Pb atoms have an apparent height of about +0.15 nm in STM, and when the junction impedance is reduced to 120 k$\Omega$, a highly corrugated type of pulling manipulation is reported, that matches the surface lattice constant.  Again using Eq.~(\ref{DeltamuSTM2}), an attractive potential well 260 meV deep is obtained.  When the tip is moved even closer to the Pb atom (43 k$\Omega$) Bartels \textit{et al.} report a much smoother manipulation process which they describe as a sliding motion.  In this case I predict an induction well 730 meV deep.  Such a deep well should trap a Pb atom with high efficiency, and this is consistent with the observation that the surface crystal corrugation has little effect at these settings.   
 
\begin{figure}[ht]
	\centering
		\includegraphics[width=0.25\textwidth]{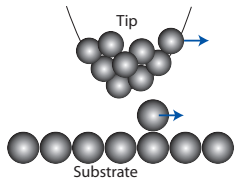}
			\caption{Schematic diagram illustrating the repulsive ``pushing'' mechanism, where the STM tip being moved to the right, causes an adsorbate to move along the same direction while located (along a step edge not shown) in front of the tip.}
		\label{fig:push}
\end{figure}

In all of the examples cited so far, the adsorbates that are amenable to manipulations image as bright features in STM.  In sharp contrast, CO appears as a sombrero-shaped depression when imaged at 0.05 V on the Cu(211) surface~\cite{Meyer97}.  The apparent height in this case is $a\approx -0.06~$nm.  Bartels \emph{et al.} have shown that pulling and sliding type manipulations fail for this system but a repulsive manipulation (pushing) does succeed, at a junction impedance of 390 k$\Omega$.  This pushing mechanism succeeds because the Cu(211) surface has high anisotropy.  The CO molecules adsorb onto step edges and they can be pushed along the step edges without being quickly lost.  See Fig.~\ref{fig:push} for a depiction of this process.  Here, the tip moves to the right and pushes the CO adsorbate while the CO is in front of the tip and is never trapped by the tip.  This is the only study that I am aware of that unambiguously demonstrates a repulsive manipulation.  I think it is highly significant that this coincides with a negative value for the apparent height.  Thermodynamic induction explains this very nicely.  With $a<0$ the induced potential becomes repulsive.  My calculations give a repulsive barrier height of 32 meV.  This makes pushing possible, to a certain extent, but not trapping.   
I believe that this example represents a major success for my model and begins to place thermodynamic induction on sound footing.  My reasoning is that though models for interactions between tip and adsorbates exist (see~\cite{StroscioCelotta91,Kantorovich2008} for examples) there is no known force model that predicts CO-tip attraction on one metal surface and CO-tip repulsion on a very similar surface.
The argument for TI becomes a compelling one. 

\subsubsection{Further proposals for STM-based tests for TI}

\begin{figure}[ht]
	\centering
		\includegraphics[width=0.25\textwidth]{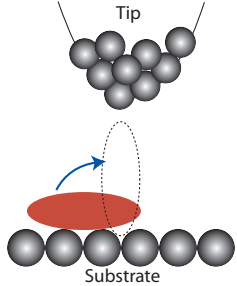}
		\caption{Schematic diagram showing how TI can lead to the rotation of a molecule into a configuration that increases the tunnel current.  The $\Delta F_{G,qss}$ free energy budget could go towards aiding the transition to a higher energy configuration.}
	\label{fig:Rotate}
\end{figure}

\begin{figure}[ht]
	\centering
		\includegraphics[width=0.25\textwidth]{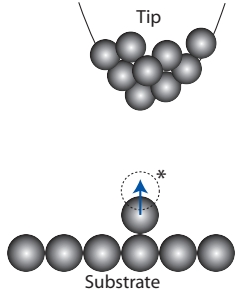}
				\caption{Schematic diagram showing how TI can guide the excitation of a resonant atomic or molecular excitation which leads to a greater junction conductance via resonant tunneling.}
			\label{fig:Excited}	
\end{figure}

\begin{figure}[ht]
	\centering
	\includegraphics[width=0.25\textwidth]{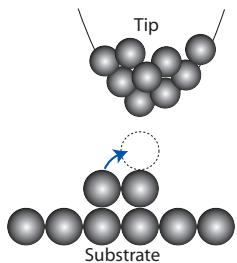}
\caption{Schematic diagram showing the STM induced on-top trapping process.  This diffusion event involves both lateral motion and motion perpendicular to the surface.}
			\label{fig:Ontop}
\end{figure}

In Figs.~\ref{fig:Rotate}, \ref{fig:Excited}, and \ref{fig:Ontop} I indicate schematically three further possible scenarios, where TI is predicted to have observable effects by creating potential wells suitable for trapping atoms or molecules into certain states.  I emphasize that the induction principles presented here must work in concert with established microscopic bond-breaking mechanisms and can lead to the possibility of huge enhancements in the ease of these manipulations only if the tunnel junction conductance is significantly increased.   For the first two schemes, careful analysis might be able to distinguish TI effects from competing explanations such as electric field-assisted diffusion and traditional electromigration~\cite{Eigler91}.  For the third scheme, it is less likely that such traditional mechanisms alone would provide the definitive explanation, i.e., success with the third scheme would be a major step in solidifying the framework of TI.  

In Fig.~\ref{fig:Rotate} I propose that TI would lead to rotation and trapping of a molecule that is elongated in shape.  After exciting a rotation transition, the tunnel junction would have an increased conductance.  This would require a molecule with electronic structure that has a through-molecule tunneling rate greater than that of the same length of vacuum.  Molecular chains with small band gaps would make good candidates~\cite{Fisher1999}.  Potentially, a long molecule could give a large effective value for $a$, which would give a deep TI potential well.   

For the case shown in Fig.~\ref{fig:Excited}, an adsorbate is promoted to a state resembling a transition state.  An example is a temporary negative ion resonance, such as the case for CO molecules on Cu(111) excited to a $2\pi^*$ resonant state and studied by both STM and time-resolved two-photon photoemission~\cite{Meyer1998}.  If the excitation leads to a higher junction conductance, via the electronic resonance, then TI is predicted to have an effect on the outcome.  In this case, the reported outcome was efficient transfer of CO from the surface to the end of the STM tip.  I propose that while in the resonant state, the CO species moves on a different potential energy curve, i.e., the excited state potential energy curve, and that this likely means the center-of-mass of the CO lies further from the surface than when in the ground state (see dashed circle in Fig.~\ref{fig:Excited}).  In this case, I predict that TI would induce the CO outwards away from the surface (on average), and would facilitate the jump to the tip.  The induced $\Delta\mu_{G,qss}$ value would be negative and would lower the energy barrier for the jump.  I also point out that in this case there would be no problem getting the gate subsystem into the quasistationary state since the species is excited and positions of the C and O nuclei would be expected to fluctuate significantly and quickly on the subpicosecond timescale.

Numerical simulations of the dislodging of chemisorbed benzene from Si(100) via negative ion resonance shows that significant fluctuations do indeed occur on the 0.1 ps timescale and that the excited-state potential does push the molecule center-of-mass away from the surface~\cite{PatitsasPRL2000}.  Though the covalent bonding of benzene/Si means the energy barrier for desorption is higher ($\approx$ 1 eV) the STM-induced dislodging occurs at higher applied bias, in the range of 2-3 V~\cite{Patitsas2000}.  Unfortunately the conductance of the junction during the short-lived ion resonance is not known.  I merely note that for resonant tunneling in general, the junction conductance could theoretically increase dramatically.  As an instructive example calculation, if the momentary conductance increases to $G=10^{-6}$ $\Omega^{-1}$, then $\dot{F}_{extra}=-G W^2\approx 4\times 10^{-6}$ W and by Eq.~(\ref{kquartic2}) $\Delta F_G=1.3~$eV.  This energy budget exceeds the binding energy of 1.05 eV for benzene/Si(100) while in the chemisorbed state just preceding the STM-induced dislodging~\cite{Patitsas2000,Lopinski1998}. Though this budget could be used to overcome the barrier, I offer no detailed explanation for just how this would happen.  While the conductance value used was very large, it is certainly reasonable, and it is still substantially smaller than the (conductance quantum) theoretical limit for resonant tunneling, $G_Q=e^2/\pi\hbar=7.75\times 10^{-5}$~$\Omega^{-1}$~\cite{Landauer1957}.   Though a thorough analysis awaits further knowledge of the resonant conductance, this simple calculation does support the idea that TI could play an important role in the manipulation of chemisorbed species, at least at higher applied bias.        

The scheme shown in Fig.~\ref{fig:Ontop} is similar to Fig.~\ref{fig:push}  though the induced diffusion is both lateral and normal to the surface.  The promotion of the candidate atom to the on-top configuration (dashed outline) will likely require an energy investment, similar to the resonant excitation just discussed, and will likely be more difficult to complete than the simple lateral diffusion.  Inspection of Eq.~(\ref{Var2SSc3}) suggests that by holding the tip-sample distance constant (so $G$ does not change) while increasing the bias $W$, the success of this approach should be made more likely, since the induced potential well will be deeper.  Evidence for success would be a sudden upwards jump in the tunnel current.  Also, from the trapped on-top configuration, the atom may make a subsequent transition to the end of the tip.  After the experiment, subsequent imaging scans showing the atom residing nearby on the surface would supply further evidence that it was moved.

\subsubsection{Induction-stabilized atomic tether}


\begin{figure}[ht]
	\centering
		\includegraphics[width=0.25\textwidth]{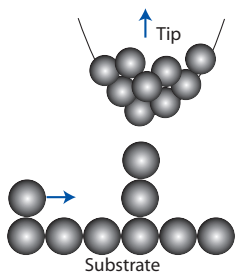}
		\caption{Illustration showing a third atom diffusing towards the tunnel junction after a successful on-top result.  Also shown is the user-controlled tip motion away from the surface, to accommodate the incoming atom.}
		\label{fig:3rdComing}
\end{figure} 

\begin{figure}[ht]
	\centering
		\includegraphics[width=0.25\textwidth]{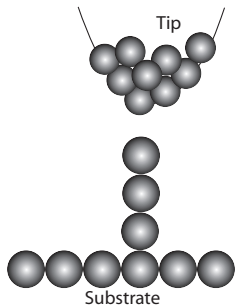}
		\caption{Illustration of the three atom tether.  The tunneling conductance of the tether junction must be appreciably higher than that of vacuum.}
		\label{fig:3atomtether}
\end{figure}  

Though the possibilities for using TI to manipulate atoms are tantalizing and the numerical calculations presented here strongly support TI, conclusive evidence for TI remains to be found (though the CO repulsion comes close).  Other explanations have been proposed and I admit that it is almost impossible to create a real scenario in STM that does not have any of the following effects: ionization by strong electric fields, vibrational excitation with local heating, electronic excitation, and transient ionic resonance effects.  The scenarios shown in Figs.~\ref{fig:Rotate}, \ref{fig:Excited}, and \ref{fig:Ontop}, even if successful, could all in principle, be attributed to these accepted mechanisms.

In this subsection I discuss a scheme that if attempted and verified, would provide conclusive evidence for TI, since all other explanations are easily ruled out.  The tether scheme begins with the positive outcome of the on-top diffusion experiment, shown in Fig.~\ref{fig:Ontop}.  As I have pointed out, as is, this would be a remarkable outcome.  Getting a third atom to sit on top of the first two would be considered impossible by accepted mechanisms, as the energetics become prohibitive.  For thermodynamic induction, the rate of entropy increase plays an important role in $\dot{F}$ and this rate can stabilize the structure shown in Fig.~\ref{fig:3atomtether}.  In order to carry this out by STM, one must have the correct procedure.  Immediately after achieving the on-top outcome of Fig.~\ref{fig:Ontop} the junction conductance would be high, and there may not be enough physical space to accommodate one more atom.  From the TI perspective, adding another atom may not increase the junction conductance, so the driving force towards a 3-atom tether would not exist.  If, however, the tip is moved upwards by an amount comparable to atomic dimensions (perhaps 0.2 nm or so) then the situation changes drastically.  The junction impedance would rise (by roughly a factor of 100) and also the physical space is made available for one more atom under the tip.  Since the addition of this third atom, after the tip motion, would significantly increase $G$, the inductive driving force would be strong.  This step is illustrated in Fig.~\ref{fig:3rdComing}, with the upwards-pointing arrow depicting the user-controlled tip motion, most likely carried out using piezoelectrics.  Indeed, this outwards tip motion may be necessary to accomplish the on-top diffusion event of Fig.~\ref{fig:Ontop}.  Both the on-top event and creation of the three-atom tether would be difficult since one or two atoms wind up in energetically unfavorable conditions.    
To overcome this the magnitude of $\Delta\mu_{G,qss}$ must be large, at least on the same scale as the adsorption energies for the tether atoms.  This will probably require operating the tunnel junction very close to the maximum electric field that can be withstood.  On this point the outward tip motion illustrated in Fig.~\ref{fig:3rdComing} can be beneficial: after the motion, the applied bias $W$ can be raised to a higher value, with the junction staying below the critical electric field.  According to Eq.~(\ref{Var2SSc3}), this will deepen the $\Delta\mu_{G,qss}$ potential well and will stabilize the tether.

This process can, in principle, be repeated indefinitely, specifically: 1) receive a new atom into the tether, 2) pull the tip outwards, 3) increase the applied bias, and 4) wait for the next atom.  This growth process involves both self-assembly (diffusion) as well as precisely controlled steps.  The process should actually get easier each time, as the number of atoms in the tether $N_{tether}$ grows, $\Delta\mu_{G,qss}$ will get larger and larger.  (Recall that $\Delta\mu_{G,qss}$ is a per particle energy.)  The key is that $\Delta\mu_{G,qss}$ in Eq.~(\ref{Var2SSc3}) depends on $W$ but not on the tip-sample separation distance.

After repeating this specialized growth process often enough, with $N_{tether}$ large enough, $\Delta\mu_{G,qss}$ will eventually get into the range of several eV.  Subsequently, the operator of the experiment can choose to stop increasing $W$ in each cycle.  In this case, there seems to be no limit to  $N_{tether}$, for example tether lengths of 1 m and longer are obtainable.  In order to keep the tether growing, one must keep supplying the junction with more adsorbates, and one must also avoid clumping, by making sure the rate at which diffusing adatoms impact the tunnel junction is not too high.  The experimenter must have sufficient time to pull the tip back after confirming the addition of the latest tether atom.  The mean time interval between surface atom impacts is $b^2/4D\theta$, where $\theta$ is the surface coverage~\cite{Zangwill}.  As long as conditions such as dosage and temperature are set correctly this time interval can be maintained to be large enough to accommodate the tip adjustment and therefore keep the tether one atom thick and ``taut''.  At any moment one could choose to cut off the supply of adsorbates by either turning the flux from the gas phase off, or by lowering the temperature and freezing out the surface diffusion.  The tether should survive as a stable structure indefinitely.  If, after creating a tether, $W$ gets stepped upwards after each cycle, eventually to values beyond 1 kV and even approaching the MV range, then $\Delta\mu_{G,qss}$ can attain values not considered attainable in any condensed matter system.  Even with a modest junction current of 1 nA, potential wells 1 MeV deep are possible when these parameters are inserted into Eq.~(\ref{DeltamuSTM2}).  Such a tether may contain a million atoms with a total length around 0.2 mm.  A cohesive energy of 1 MeV per particle is well beyond that of even the strongest materials such as diamond and steel, suggesting incredible strength and stability.  Moreover, these nuclear-level chemical potentials invite the possibility of TI catalyzing nuclear reactions within the tether.  Thermodynamic induction would always push the system towards states of higher tether conductivity.   Lastly, I consider the theoretical limits:  If the junction conductance is maintained at $G_Q$, the conductance quantum, then the chemical potential difference scales as $W^2$ and at 1 MV, $\Delta\mu_{G,qss}$ can take values in the TeV range.  Such systems would surely be of great interest to the high energy physics community.  For example, a sudden disruption of the tether may lead to an atom escaping with several TeV of kinetic energy.  

The atomic tether is a striking example of OBI, especially when significant length is achieved, for example a chain of 100 atoms strung between tip and sample, just one atom thick would clearly be a highly ordered structure.  The 100 atoms would have much less entropy while in the tether state as compared to the same 100 atoms diffusion randomly on the surface.  Self-organization is probably not an apt term since a considerable experimental procedure is required to create the structure.  The tether certainly does not resemble systems with biological complexity.  Nevertheless, I believe that this OBI structure would have significant bearing on the discussion of OTF.  I believe that the atomic tether would, in the words of Prigogine and Nicolis, constitute an example of a departure from incoherent behavior.  

\section{Continuum Diffusion Model in One or Two Dimensions}  \label{sec:cont}

So far, all the results derived have been for discrete variables.  The proposals outlined in Section~\ref{sec:examples} made extensive use of my result for the change in chemical potential as described by Eqns.~(\ref{Var2SSc},~\ref{Var2SSc2}).  In the context of the examples of Section~\ref{sec:examples} one should think of these as the net change in some continuous potential function.  Thus it is desirable to work out a scheme for calculating the shapes of these potential profiles.  In this section an extension of the standard diffusion equation is obtained.  By adding in inertial terms one could, in future work, treat fluid flow with TI effects included.  Fluid dynamics has been studied in the context of nonequilibrium thermodynamics, including precise establishment of the Onsager relations in the linear realm~\cite{McLennan1974,Dufty1987}.  The nonlinear aspects of having variable kinetic coefficients could be accounted for by following McLennan's approach which utilizes the Liouville equation, which in turn was built on the approach presented by Bernard and Callen~\cite{Callen1959,McLennan1961}.  The work presented in this section suggests a natural approach for exploring TI effects in fluids, including finding induction effects on mean values, time reversal properties and correlation functions~\cite{McLennan1968,Dufty1968,Heinz2002}.
In order to determine what happens to the thermodynamic induction terms in the continuum limit, in the context of diffusive particle transport, I first discuss the continuum limit without induction.

\subsection{Continuum limit without induction}

In order to set up the continuum limit I consider for now the case where there is just one dynamical reservoir variable (slow) $a_{slow}$, and a discrete set of fast variables $a_i$ which represent two dimensional cells laid side by side along a straight line.  Each cell is a flat square with side $w$.  Generalization of Eq.~(\ref{Lij}) to this case gives 
\begin{equation}
\dot{a}_k=\frac{D}{w^2}(a_{k+1}-a_k)-\frac{D}{w^2}(a_{k}-a_{k-1})  -\lambda_k a_k~,  \label{akDlambda}
\end{equation}
where the first two terms describe diffusion between cells and the last term represents diffusion elsewhere.  The last term could represent diffusion at the boundaries, but not necessarily.  For example, for the case of surface diffusion, this last term could describe adsorption from or desorption to the gas phase.  
If $n(x)$ represents the areal particle density deviation from the equilibrium value $n_{eq}$ then the continuum limit of Eq.~(\ref{akDlambda}) (without induction) is
\begin{equation}
\frac{\partial n}{\partial t}=D\frac{\partial^2 n}{\partial x^2} -\lambda(x) n~.  \label{RegDiffEq}
\end{equation}

Before proceeding to the induction treatment, I note that in terms of the forces, $Y_k=-a_k/c$, where I have extended Eq.~(\ref{zeta}) to $c\equiv \bar{N}_k/(\zeta k_B)$. 
If one implements the correlation functions from Eq.~(\ref{LijKij}) one may write
\begin{equation}
\dot{a}_k =L_{OD}Y_{k-1}+L_D Y_k +L_{OD}Y_{k+1}
\end{equation}
where the off-diagonal coefficients are given by
\begin{equation}
L_{OD}= -\frac{cD}{w^2} =\frac{1}{k_B}\int_0^{\infty}ds\langle\dot{a}_{k+1}(s)\dot{a}_k(0)\rangle_0   \label{LOD}
\end{equation}
and the diagonal coefficient is
\begin{equation}
L_{kk}= c\lambda_k -2L_{OD} =\frac{1}{k_B}\int_0^{\infty}ds\langle\dot{a}_k(s)\dot{a}_k(0)\rangle_0~. \label{Lkk}
\end{equation}

\subsection{Continuum limit with induction included}

Each cell will have a different conductance $G_l$ so there will be a set of Onsager coefficients $L_{Dl}$ and $M_{Dl}$ where the subscript $D$ refers to the single dynamical reservoir variable.  Each partial conductance $G_l$ is assumed to depend on $N_l$ (only).  Making use of Eqs.~(\ref{MtoG}) and (\ref{gamm112}) gives
\begin{equation}
\gamma_{D,l}=\frac{\partial M_{Dl}}{\partial N_l}=\frac{T}{e^2}\frac{\partial G_l}{\partial N_l} =\frac{T}{e^2}\xi_{l}~,  \label{gammaDl}   
\end{equation}
where I have defined $\xi_{l}\equiv \frac{\partial G_l}{\partial N_l}$.
 
Since the off-diagonal correlation functions are nonzero I derive the induction term directly.  Making use of Eq.~(\ref{akindfull}) and Eq.~(\ref{gammaDl})  gives
\begin{equation}
\left(\dot{a}_k\right)_{ind}=\frac{1}{k_B\Delta t_k}Y_{slow}^2 \frac{T}{e^2 }\sum_{l=k-1}^{k+1}\xi_l\int_{t}^{t+\Delta t_k}dt'\int_t^{t'}dt''\int_{-\infty}^{t''}dt'''\langle\dot{a}_k(t')\dot{a}_l(t''')\rangle_0~. 
\end{equation}
Making use of Eq.~(\ref{LOD}) and Eq.~(\ref{Lkk}) and closely following the steps explained in Ref.~\cite{Patitsas2014} results in
\begin{equation}
\left(\dot{a}_k\right)_{ind}= \frac{T\tau^* Y_{slow}^2}{e^2 }\left[ c\lambda_k \xi_k +L_{OD}(\xi_{k+1}-2\xi_{k}+\xi_{k-1} )\right]~. 
\end{equation}
The continuum limit is then
\begin{equation}
\left(\frac{\partial n}{\partial t}\right)_{ind}= \frac{\tau^* W^2 n_{eq}}{\zeta k_B T}\left[\lambda(x) \xi(x) -D\frac{\partial^2\xi}{\partial x^2}\right]~. \label{nind}
\end{equation}
Note, I have defined the continuum limit function $\xi(x,y)$ as 
\begin{equation}
\xi \equiv \frac{\partial g}{\partial n}~,   \label{xicontdef}
\end{equation}
where $g$ is defined as the conductance per unit area.
Comparing Eq.~(\ref{nind}) to Eq.~(\ref{RegDiffEq}) gives the generalized diffusion equation
\begin{equation}
\frac{\partial n}{\partial t}= D\nabla^2 n- D \frac{\tau^* W^2 n_{eq}}{\zeta k_B T} \nabla^2\xi -\lambda n +\lambda\frac{\tau^* W^2 n_{eq} }{\zeta k_B T}\xi ~.  \label{GenDiffEq2}
\end{equation}
Induction enters as two distinct terms in this equation, one for diffusion and the other for adsorption.

The net particle flux $\jvec_{2,net}$ is a sum of two fluxes, one the standard flux caused by diffusion, i.e., $\jvec_{diff}=-D\del n$,
and the other term $\jvec_{ind}$ deals with induction.  The induction flux is given by
\begin{equation}
\jvec_{ind}(\rvec)= D\frac{\tau^* W^2 n_{eq}}{\zeta k_B T}\del \xi ~. \label{jind}
\end{equation}
Thus, $\jvec_{net}=\jvec_{diff}+\jvec_{ind}$ and this result constitutes a generalization of Fick's law.  For small deviations from the equilibrium particle density $n_{eq}$  
\begin{equation}
\del\mu=\zeta\frac{k_B T}{n_{eq}}\del n ~. \label{delmu}
\end{equation}
In many instances the gradient of the chemical potential is equivalent to a force; for example in the analysis of thermoelectric properties of materials.  By combining Eqs.~(\ref{jind}) and (\ref{delmu}), one may identify an inductive force as
\begin{equation}
\Fvec_{ind} = \tau^* W^2 \del\xi ~.  \label{Find}
\end{equation}
Thermodynamic induction also gives rise to a per particle potential energy $U_{ind}$ related to the induced force by $\Fvec_{ind}=-\del U_{ind}$.  Thus $U_{ind}$ is given by
\begin{equation}
U_{ind}(x,y)\equiv -\Delta\mu(x,y)=-\tau^* W^2  \xi(x,y) ~.  \label{Uind}
\end{equation}
The induced force acts just like an external field such as the term $\sigma\Evec/e$ that would account for charged carrier diffusion with an external electric field.  In terms of the potential energy, $U_{ind}$ could cancel off a repulsive external potential.

\subsubsection{Quasistationary continuum states}

The continuum limit version of the quasistationary state corresponds to setting $\frac{\partial n}{\partial t}=0$ in Eq.~(\ref{GenDiffEq2}).  Under the condition that the function $\xi$ goes to zero at infinity, then $n$ must also go to zero at infinity.  Under this condition the only solution is
\begin{equation}
n_{qss}(x,y) = \frac{\tau^* W^2 n_{eq}}{\zeta k_B T}\xi(x,y) ~. \label{Dzero}
\end{equation}  
Since the change in $n$ is assumed small, then from Eq.~(\ref{delmu}) the chemical potential profile is given by 
\begin{equation}
\Delta\mu_{qss}(x,y)= \zeta\frac{k_B T}{n_{eq}}n_{qss}(x,y) = \tau^* W^2\xi(x,y)=-U_{ind}(x,y)~.  \label{Deltamuqss}
\end{equation} 

\subsection{Variational principles}

From Eq.~(\ref{akDlambda}) one obtains for the free energy production inside a discrete cell (without induction) the expression
\begin{equation}
\dot{F}_k= -T\dot{a}_k Y_k = T c^{-1} a_k\left[ \frac{D}{w^2}(a_{k+1}-a_k)-\frac{D}{w^2}(a_{k}-a_{k-1})  -\lambda_k a_k \right]~.   \label{FkDlambda}
\end{equation}
In order to take the continuum limit I define the density of free energy production as
\begin{equation}
\dot{f}_k \equiv \frac{\dot{F}_k}{w^2}~. \label{fkdef}
\end{equation}
In taking the continuum limit, while including the induction terms from Eq.~(\ref{GenDiffEq2}), I split the fast term into a contribution from diffusion and a contribution from adsorption/desorption, to give
\begin{equation}
\frac{\partial {f}_{diff}(x,y,t)}{\partial t}=\frac{\zeta k_B T}{\bar{n}}n\left(D\nabla^2 n-D\frac{\tau^* W^2 n_{eq}}{\zeta k_B T}\nabla^2\xi \right)~,   \label{dotfDiff}
\end{equation}
\begin{equation}
\frac{\partial {f}_{ads}(x,y,t)}{\partial t}=-\frac{\zeta k_B T}{\bar{n}}n \lambda\left[n -\frac{\tau^* W^2 n_{eq}}{\zeta k_B T}\xi \right] ~.  \label{dotfAds}
\end{equation}
Specifically for system A, I define the following thermodynamic power functions as 
\begin{equation}
\dot{F}_{diff}\equiv\int \frac{\partial {f}_{diff}}{\partial t} dxdy~,
\end{equation}
\begin{equation}
\dot{F}_{ads}\equiv\int \frac{\partial {f}_{ads}}{\partial t} dxdy~,
\end{equation}
\begin{equation}
\dot{F}_{slow}\equiv\int \frac{\partial {f}_{slow}}{\partial t} dxdy = -W^2\int g(n) dxdy~.
\end{equation}
For the rest of this subsection I take $D=0$, and leave the $D\neq 0$ case for later study.  In this special case $\dot{F}_{fast}=\dot{F}_{ads}$ and the timescale $\tau_{ads}=\lambda^{-1}$. In this case I formulate the important combination
\begin{equation}
\Psi\equiv {F}_{slow}- r_{ads}^{-1} F_{ads}~,
\end{equation}
where $r_{ads}\equiv \tau^*/\tau_{ads}$.  Similarly to Eqs.~(\ref{PhiDef}) and (\ref{PhiDotDef}) have made use of the time derivative:
\begin{equation}
\dot{\Psi}\equiv\dot{F}_{slow}- r_{ads}^{-1}\dot{F}_{ads}~.
\end{equation}

The direct link between the slow and fast subsystems is made by assuming that the (slow variable) conductance $g$ depends on $n$.  As an integral, $\dot{\Psi}=\int \psi(n, n_x, n_y) dxdy$~,
where $n_x$ is short-form for $\frac{\partial n}{\partial x}$, etc. and 
\begin{equation}
\psi=-W^2 g(n) + \frac{r_{ads}^{-1}\zeta k_B T\lambda}{n_{eq}}n\left(n -\frac{\tau^* W^2 n_{eq}}{\zeta k_B T}\xi\right)~. \label{psidef}
\end{equation}
If this functional is minimized by setting $\delta\dot{\Psi}=0$ then the Euler-Lagrange equations give
\begin{equation}
\frac{\partial\psi}{\partial n}-\frac{\partial}{\partial x}\frac{\partial\psi}{\partial n_x}-\frac{\partial}{\partial y}\frac{\partial\psi}{\partial n_y}-\frac{\partial}{\partial z}\frac{\partial\psi}{\partial n_z}=0~, \label{ELeqn}
\end{equation}
or
\begin{equation}
-W^2\xi  + \frac{r_{ads}^{-1}\zeta k_B T\lambda}{n_{eq}}\left(2n -\frac{\tau^* W^2 n_{eq}}{\zeta k_B T}\xi\right) =0~.
\end{equation}
After some algebra one finds the solution for $n$ 
which, as expected, is the quasistationary result Eq.~(\ref{Dzero}).  Thus I have obtained the continuum version of Theorem 2; the quasistationary continuum state minimizes $\dot{\Psi}$.
One is also left with a very good approach for extending to the case where $g(n)$ is not necessarily a linear function.  This approach would require solving a nonlinear partial differential equation.  If  Eq.~(\ref{Dzero}) is substituted back into Eq.~(\ref{psidef}) then one finds that
\begin{equation}
\dot{F}_{extra}= -\frac{\tau^* W^4  n_{eq} }{\zeta k_B T}\int \xi^2 dxdy~.   \label{dotFcont}
\end{equation}
That this quantity can never be positive verifies the continuum version of Theorem 1.  In the quasistationary continuum state, the DR always approaches equilibrium faster as compared to when $n=0$.

\subsection{Induction potential profile for STM tunnel junction}

Towards obtaining an induced potential energy profile for diffusing surface adsorbates near an STM tunnel junction, I invoke the Tersoff--Hamann approach for the tunnel current~\cite{Tersoff1985,Patitsas2009}.  The result is the form
\begin{equation}
\xi=\xi_0 e^{\alpha(-L+R)}~,
\end{equation}
where $L=\sqrt{R^2+x^2}$ and $x$ is the profile parameter with the position directly under the tip designated as $x=0$.  The variable $x$ runs along the surface.  Applying Eqs.~(\ref{DeltamuSTM2}) and (\ref{Uind}) the potential energy profile is
\begin{equation}
U(x)= -\alpha a U^* e^{\alpha(-L+R)}~.
\end{equation}
Note that when $x=0$ one recovers the correct well depth $U_{fast,qss}$.  Since STM achieves high spatial resolution, with $R$ being on the order of a nanometer or even less, then $U(x)$ will also be highly localized.  The force profile is readily obtained as
\begin{equation}
F(x)=-\frac{d}{dx}U(x)= -\alpha^2 a U^* e^{\alpha(-L+R)} \frac{x}{\sqrt{R^2+x^2}}~.
\end{equation}
For small $x$ the force constant is 
\begin{equation}
k= \frac{\alpha^2 a U^*}{R}= \frac{\alpha \Delta\mu_{G,qss}}{R}~,
\end{equation}
where I made use of Eq.~(\ref{DeltamuSTM2}).  Putting in numbers for the case of CO/Pt(111), $\Delta\mu_{G,qss}=30~$ meV, $\alpha=22$ nm$^{-1}$, $R=0.7~$nm I obtain $k=0.15$~N/m.  This number is comparable to lateral force constants of cantilevers typically used in atomic force microscopy \cite{MeyerAFMbook}.  I conclude that some type of scheme using a combination STM/AFM could be used to measure this lateral thermodynamically induced force. Thus I have derived both the depth and shape of the potential well for an adsorbate diffusing over a surface near the STM tunnel junction.  I emphasize that these key considerations were entropic in nature so the result is an entropic trap for the adsorbate located directly beneath the STM tip.  This is a new type of entropic trapping complements the sort of trap used for macromolecules~\cite{Asher1999,Craighead1999} with the distinguishing feature being that for the STM trap, there are two thermodynamic variables involved, with the trapped particle acting as a gate for the DR.

\section{Conclusions}

Following up on the establishment of TI in Ref~\cite{Patitsas2014}, I have developed a theory for TI under isothermal conditions.  In my approach thermodynamic variables are not directly coupled energetically, but instead they are coupled in the term describing the rate of entropy production.  Naturally this implies coupling within the terms describing the rate change of the Helmholtz free energy.
In addition I have shown how TI is treated in the continuum limit.
A clear distinction is made between dynamical reservoir variables and faster, gate variables.  If one forgets for a moment the reservoir variables, then it would appear that the gate variables are spontaneously pushed well away from equilibrium for sustained periods of time.  Seemingly, the second law of thermodynamics would be violated.  Of course, no such violation takes place since the reservoir variables must be accounted for and when they are, the total system entropy never decreases.

The analysis naturally leads to thermodynamic functions that are the rate of change in time of thermodynamic free energy functions.  The stationary states introduced by Prigogine play an important role in understanding the meaning of the induction terms.  An important theorem concerning the free power $\dot{F}_{extra}$ states that the Helmholtz free energy always approaches equilibrium faster when gate variables are quasistationary.  This holds true in both the discrete and continuum cases.  Establishing these quasistationary states requires energy and I have formulated an expression for precisely how much of a free energy budget is available.  This energy budget could be used to help create interesting physical structures.  A variational principle I have presented here works for minimizing the potential rate function, $\dot{\Phi}$, and similarly for the continuum limit.  This variational approach allows for a straightforward approach towards dealing with highly nonlinear systems.  I believe the general results presented here can add significantly to the understanding of entropic trapping as well as to OBD effects in frustrated spin systems. 

Several important examples of isothermal TI have been discussed, including a type of transverse electromigration/electro-osmosis that may be detectable in electrolytes and superfluid helium containing ions.  As applications, TI could be used to analytically purify solvents or used to drive ionic fluid motion.  In further technologically important examples I show that TI can lead to significant motion of carriers in devices such as field effect transistors.  Thermodynamically induced diffusion of dopants may play an important role in semiconductor device failure. 

As an example of a bottlenecked system exhibiting enhanced TI, STM-based manipulation and entropic trapping of atoms and molecules has been discussed in detail.  Several schemes designed for isolating TI effects have been outlined.  A general rule of thumb emerges from the analysis: adsorbates diffusing around on the surface will be induced to either spend more time or less time directly in the tunnel junction so that in either case the junction conductance is increased.  For example, a potential well for such an adsorbate can be completely entropic in origin, can have depth comparable to $k_B T$ and would be additive to any other potentials that are strictly energy based.  Also, a thorough review of previously reported manipulations, supported by calculations for potential wells or barriers, provides strong support for the idea that accepted microscopic bond-breaking mechanisms are governed by a general thermodynamic principle.

In particular, I believe that TI is necessary to explain the level of control that sliding-type manipulations demonstrate.  The most reasonable choices for the parameters input into the simple formula gives a threshold condition for STM manipulations that is strikingly close to what is required to match results reported in the literature.  This is not to be taken lightly as the only adjustable parameter in the calculation is the pre-exponential factor which cannot differ from $10^{13}~$s$^{-1}$ by very much.  It is highly unlikely that one can attribute this to coincidence, especially since the windows for operating parameters (applied bias and tunnel current) are quite narrow.  In a sense, STM barely works as a surface characterization technique since rather small changes away from settings that give successful characterization quickly lead to either poor levels of usable signal or to the threshold for disruption of the surface.    Also, I present a compelling argument that only TI can explain the pushing-type manipulation results for CO/Cu(211).   My calculations provide hints that TI could play an important role in catalytic activity.   My continuum model predicts the shape of the adsorbate potential well for the STM case and my calculations for the lateral force on just one atom predict that a substantial force should be detectable by an atomic force microscope.  I hope that improved understanding of TI will aid in developing future technologies that involve precise atom-by-atom manipulations.  

All of the realistic scenarios described in this work are contained inside the VKC class of systems.  I suggest that the VKC class is quite general and I believe that many more VKC systems will be discovered soon.  I propose future STM-based experiments on VKC systems such as controlled rotation of molecules, on-top diffusion events, and excited resonance-state events.  Always, the guiding principle is that the gate will act in a way to allow the dynamical reservoir to produce entropy at a greater rate.  My most remarkable proposal is for a tether between sample and tip that is just one atom thick.  This proposal involves a sophisticated technically assisted form of order-by-induction.  I have included a careful description of how such a tether could be constructed. 
	
\section{Acknowledgments}

I thank Cathy J. Meyer for her assistance in editing the manuscript.

\bibliography{EntMan31Mar2015}

\providecommand{\noopsort}[1]{}\providecommand{\singleletter}[1]{#1}%
\begin{thebibliography}{67}
\providecommand{\natexlab}[1]{#1}
\providecommand{\url}[1]{\texttt{#1}}
\providecommand{\urlprefix}{URL }
\expandafter\ifx\csname urlstyle\endcsname\relax
  \providecommand{\doi}[1]{doi:\discretionary{}{}{}#1}\else
  \providecommand{\doi}[1]{doi:\discretionary{}{}{}\begingroup
  \urlstyle{rm}\url{#1}\endgroup}\fi
\providecommand{\bibinfo}[2]{#2}

\bibitem[{Onsager(1931{\natexlab{a}})}]{Onsager1931}
\bibinfo{author}{L.~Onsager}, \bibinfo{title}{Reciprocal Relations in
  Thermodynamic Processes. I.}, \bibinfo{journal}{Phys. Rev.}
  \bibinfo{volume}{37} (\bibinfo{year}{1931}{\natexlab{a}})
  \bibinfo{pages}{405--426}.

\bibitem[{Onsager(1931{\natexlab{b}})}]{Onsager1931b}
\bibinfo{author}{L.~Onsager}, \bibinfo{title}{Reciprocal Relations in
  Thermodynamic Processes. II.}, \bibinfo{journal}{Phys. Rev.}
  \bibinfo{volume}{38} (\bibinfo{year}{1931}{\natexlab{b}})
  \bibinfo{pages}{2265--2279}.

\bibitem[{Patitsas(2014)}]{Patitsas2014}
\bibinfo{author}{S.~N. Patitsas}, \bibinfo{title}{Thermodynamic Induction
  Effects Exhibited in Nonequilibrium Systems with Variable Kinetic
  Coefficients}, \bibinfo{journal}{Physical Review E}
  \bibinfo{volume}{89}~(\bibinfo{number}{1}) (\bibinfo{year}{2014})
  \bibinfo{pages}{012108}.

\bibitem[{Miller(1960)}]{Miller1960}
\bibinfo{author}{D.~Miller}, \bibinfo{title}{THERMODYNAMICS OF IRREVERSIBLE
  PROCESSES - THE EXPERIMENTAL VERIFICATION OF THE ONSAGER RECIPROCAL
  RELATIONS}, \bibinfo{journal}{Chemical Reviews} \bibinfo{volume}{60}
  (\bibinfo{year}{1960}) \bibinfo{pages}{15}.

\bibitem[{Prigogine(1965)}]{Prigogine1965}
\bibinfo{author}{I.~Prigogine}, \bibinfo{title}{Steady States and Entropy
  Production}, \bibinfo{journal}{Physica} \bibinfo{volume}{31}
  (\bibinfo{year}{1965}) \bibinfo{pages}{719}.

\bibitem[{de~Groot and P.(1984)}]{mazur}
\bibinfo{author}{S.~R. de~Groot}, \bibinfo{author}{M.~P.},
  \bibinfo{title}{Non-Equilibrium Thermodynamics}, \bibinfo{publisher}{Dover
  Publications}, \bibinfo{address}{New York}, \bibinfo{year}{1984}.

\bibitem[{Zwanzig(1992{\natexlab{a}})}]{Zwanzig1992a}
\bibinfo{author}{R.~Zwanzig}, \bibinfo{title}{Diffusion Past an Entropy
  Barrier}, \bibinfo{journal}{J. Phys. Chem.} \bibinfo{volume}{96}
  (\bibinfo{year}{1992}{\natexlab{a}}) \bibinfo{pages}{3926--3930}.

\bibitem[{Zwanzig(1992{\natexlab{b}})}]{Zwanzig1992b}
\bibinfo{author}{R.~Zwanzig}, \bibinfo{title}{Dynamical disorder: Passage
  through a fluctuating bottleneck}, \bibinfo{journal}{J. Chem. Phys.}
  \bibinfo{volume}{97} (\bibinfo{year}{1992}{\natexlab{b}})
  \bibinfo{pages}{3587}.

\bibitem[{Reguera and Rub\'\i(2001)}]{PhysRevE.64.061106}
\bibinfo{author}{D.~Reguera}, \bibinfo{author}{J.~M. Rub\'\i},
  \bibinfo{title}{Kinetic equations for diffusion in the presence of entropic
  barriers}, \bibinfo{journal}{Phys. Rev. E} \bibinfo{volume}{64}
  (\bibinfo{year}{2001}) \bibinfo{pages}{061106},
  \doi{\bibinfo{doi}{10.1103/PhysRevE.64.061106}}.

\bibitem[{Reguera et~al.(2006)Reguera, Schmid, Burada, Rub\'\i, Reimann, and
  H\"anggi}]{PhysRevLett.96.130603}
\bibinfo{author}{D.~Reguera}, \bibinfo{author}{G.~Schmid},
  \bibinfo{author}{P.~S. Burada}, \bibinfo{author}{J.~M. Rub\'\i},
  \bibinfo{author}{P.~Reimann}, \bibinfo{author}{P.~H\"anggi},
  \bibinfo{title}{Entropic Transport: Kinetics, Scaling, and Control
  Mechanisms}, \bibinfo{journal}{Phys. Rev. Lett.} \bibinfo{volume}{96}
  (\bibinfo{year}{2006}) \bibinfo{pages}{130603},
  \doi{\bibinfo{doi}{10.1103/PhysRevLett.96.130603}}.

\bibitem[{Liu et~al.(1999)Liu, Li, and Asher}]{Asher1999}
\bibinfo{author}{L.~Liu}, \bibinfo{author}{P.~Li}, \bibinfo{author}{S.~A.
  Asher}, \bibinfo{journal}{Nat.} \bibinfo{volume}{397} (\bibinfo{year}{1999})
  \bibinfo{pages}{141}.

\bibitem[{Han et~al.(1999)Han, Turner, and Craighead}]{Craighead1999}
\bibinfo{author}{J.~Han}, \bibinfo{author}{S.~W. Turner},
  \bibinfo{author}{H.~G. Craighead}, \bibinfo{title}{Entropic trapping and
  escape of long DNA molecules at submicron size constriction},
  \bibinfo{journal}{Phys. Rev. Lett.} \bibinfo{volume}{83}
  (\bibinfo{year}{1999}) \bibinfo{pages}{1688}.

\bibitem[{Stroscio and Eigler(1991)}]{Eigler91}
\bibinfo{author}{J.~A. Stroscio}, \bibinfo{author}{D.~M. Eigler},
  \bibinfo{title}{Atomic and Molecular Manipulation with the Scanning Tunneling
  Microscope}, \bibinfo{journal}{Science} \bibinfo{volume}{254}
  (\bibinfo{year}{1991}) \bibinfo{pages}{1319}.

\bibitem[{Avouris and \emph{et. al.}(1996)}]{AvourisLyding96}
\bibinfo{author}{P.~Avouris}, \bibinfo{author}{\emph{et. al.}},
  \bibinfo{title}{Breaking individual chemical bonds via STM-induced
  excitations}, \bibinfo{journal}{Surf. Sci.} \bibinfo{volume}{363}
  (\bibinfo{year}{1996}) \bibinfo{pages}{368}.

\bibitem[{Stipe et~al.(1997)Stipe, Rezaei, Ho, Gao, Persson, and
  Lundqvist}]{HoPersson97}
\bibinfo{author}{B.~C. Stipe}, \bibinfo{author}{M.~A. Rezaei},
  \bibinfo{author}{W.~Ho}, \bibinfo{author}{S.~Gao},
  \bibinfo{author}{M.~Persson}, \bibinfo{author}{B.~I. Lundqvist},
  \bibinfo{title}{Single-Molecule Dissociation by Tunneling Electrons},
  \bibinfo{journal}{Phys. Rev. Lett.} \bibinfo{volume}{78}
  (\bibinfo{year}{1997}) \bibinfo{pages}{4410}.

\bibitem[{Persson and Avouris(1997)}]{AvourisPersson97}
\bibinfo{author}{B.~Persson}, \bibinfo{author}{P.~Avouris},
  \bibinfo{title}{Local bond breaking via STM-induced excitations: the role of
  temperature}, \bibinfo{journal}{Surf. Sci.} \bibinfo{volume}{390}
  (\bibinfo{year}{1997}) \bibinfo{pages}{45--54}.

\bibitem[{Bartels and \emph{et. al.}(1998)}]{Meyer1998}
\bibinfo{author}{L.~Bartels}, \bibinfo{author}{\emph{et. al.}},
  \bibinfo{title}{Dynamics of Electron-Induced Manipulation of Individual CO
  Molecules on Cu(111)}, \bibinfo{journal}{Phys. Rev. Lett.}
  \bibinfo{volume}{80} (\bibinfo{year}{1998}) \bibinfo{pages}{2004}.

\bibitem[{Patitsas et~al.(2000)Patitsas, Lopinski, Hulko, Moffatt, and
  Wolkow}]{Patitsas2000}
\bibinfo{author}{S.~Patitsas}, \bibinfo{author}{G.~P. Lopinski},
  \bibinfo{author}{O.~Hulko}, \bibinfo{author}{D.~J. Moffatt},
  \bibinfo{author}{R.~A. Wolkow}, \bibinfo{title}{Current-induced organic
  molecule-silicon bond breaking: consequences for molecular devices},
  \bibinfo{journal}{Surf. Sci. Lett.} \bibinfo{volume}{457}
  (\bibinfo{year}{2000}) \bibinfo{pages}{L425--L431}.

\bibitem[{Alavi et~al.(2000)Alavi, Patitsas, Lopinski, Wolkow, and
  Seideman}]{PatitsasPRL2000}
\bibinfo{author}{S.~Alavi}, \bibinfo{author}{S.~N. Patitsas},
  \bibinfo{author}{G.~P. Lopinski}, \bibinfo{author}{R.~A. Wolkow},
  \bibinfo{author}{T.~Seideman}, \bibinfo{title}{Inducing Desorption of Organic
  Molecules with a Scanning Tunneling Microscope: Theory and Experiments},
  \bibinfo{journal}{Phys. Rev. Lett.} \bibinfo{volume}{85}
  (\bibinfo{year}{2000}) \bibinfo{pages}{5327}.

\bibitem[{Lu et~al.(1999)Lu, Polanyi, and Rogers}]{Polanyi1999}
\bibinfo{author}{P.~H. Lu}, \bibinfo{author}{J.~C. Polanyi},
  \bibinfo{author}{D.~Rogers}, \bibinfo{journal}{J. Chem. Phys. Comm.}
  \bibinfo{volume}{111} (\bibinfo{year}{1999}) \bibinfo{pages}{9905}.

\bibitem[{Sloan and Palmer(2005)}]{Palmer2005}
\bibinfo{author}{P.~A. Sloan}, \bibinfo{author}{R.~E. Palmer},
  \bibinfo{title}{Two-electron dissociation of single molecules by atomic
  manipulation at room temperature}, \bibinfo{journal}{Nature}
  \bibinfo{volume}{434} (\bibinfo{year}{2005}) \bibinfo{pages}{367}.

\bibitem[{Hla et~al.(2000)Hla, Bartels, Meyer, and Rieder}]{Meyer2000}
\bibinfo{author}{S.-W. Hla}, \bibinfo{author}{L.~Bartels},
  \bibinfo{author}{G.~Meyer}, \bibinfo{author}{K.-H. Rieder},
  \bibinfo{title}{Inducing All Steps of a Chemical Reaction with the Scanning
  Tunneling Microscope Tip: Towards Single Molecule Engineering},
  \bibinfo{journal}{Phys. Rev. Lett.} \bibinfo{volume}{85}
  (\bibinfo{year}{2000}) \bibinfo{pages}{2777--2780},
  \doi{\bibinfo{doi}{10.1103/PhysRevLett.85.2777}}.

\bibitem[{Moresco et~al.(2001)Moresco, Meyer, Rieder, Tang, Gourdon, and
  Joachim}]{Meyer2001}
\bibinfo{author}{F.~Moresco}, \bibinfo{author}{G.~Meyer},
  \bibinfo{author}{K.-H. Rieder}, \bibinfo{author}{H.~Tang},
  \bibinfo{author}{A.~Gourdon}, \bibinfo{author}{C.~Joachim},
  \bibinfo{title}{Recording Intramolecular Mechanics during the Manipulation of
  a Large Molecule}, \bibinfo{journal}{Phys. Rev. Lett.} \bibinfo{volume}{87}
  (\bibinfo{year}{2001}) \bibinfo{pages}{088302}.

\bibitem[{Maraghechi et~al.(2007)Maraghechi, Horn, and Patitsas}]{Patitsas2007}
\bibinfo{author}{P.~Maraghechi}, \bibinfo{author}{S.~A. Horn},
  \bibinfo{author}{S.~N. Patitsas}, \bibinfo{title}{Site Selective Atomic
  Chlorine Adsorption on the Si(111)7x7 Surface}, \bibinfo{journal}{Surface
  Science Letters} \bibinfo{volume}{601} (\bibinfo{year}{2007})
  \bibinfo{pages}{L1--L5}.

\bibitem[{Sloan et~al.(2010)Sloan, Sakulsermsuk, and Palmer}]{Palmer2010}
\bibinfo{author}{P.~A. Sloan}, \bibinfo{author}{S.~Sakulsermsuk},
  \bibinfo{author}{R.~E. Palmer}, \bibinfo{title}{Nonlocal Desorption of
  Chlorobenzene Molecules from the
  $\mathrm{Si}(111)\mathrm{\text{-}}(7\ifmmode\times\else\texttimes\fi{}7)$
  Surface by Charge Injection from the Tip of a Scanning Tunneling Microscope:
  Remote Control of Atomic Manipulation}, \bibinfo{journal}{Phys. Rev. Lett.}
  \bibinfo{volume}{105} (\bibinfo{year}{2010}) \bibinfo{pages}{048301},
  \doi{\bibinfo{doi}{10.1103/PhysRevLett.105.048301}}.

\bibitem[{Bernard and Callen(1959)}]{Callen1959}
\bibinfo{author}{W.~Bernard}, \bibinfo{author}{H.~B. Callen},
  \bibinfo{title}{Irreversible Thermodynamics of Nonlinear Processes and Noise
  in Driven Systems}, \bibinfo{journal}{Rev. Mod. Phys.} \bibinfo{volume}{31}
  (\bibinfo{year}{1959}) \bibinfo{pages}{1017--1043}.

\bibitem[{Reif(1965)}]{Reif15}
\bibinfo{author}{F.~Reif}, \bibinfo{title}{Fundamentals of Statistical and
  Thermal Physics}, chap.~\bibinfo{chapter}{15}, \bibinfo{publisher}{McGraw
  Hill}, \bibinfo{address}{New York}, \bibinfo{year}{1965}.

\bibitem[{Kampen(1985)}]{VanKampen1985}
\bibinfo{author}{N.~V. Kampen}, \bibinfo{title}{Elimination of fast variables},
  \bibinfo{journal}{Physics Reports}
  \bibinfo{volume}{124}~(\bibinfo{number}{2}) (\bibinfo{year}{1985})
  \bibinfo{pages}{69 -- 160}, ISSN \bibinfo{issn}{0370-1573},
  \doi{\bibinfo{doi}{http://dx.doi.org/10.1016/0370-1573(85)90002-X}}.

\bibitem[{Prigogine(1967)}]{Prigogine}
\bibinfo{author}{I.~Prigogine}, \bibinfo{title}{Introduction to Thermodynamics
  of Irreversible Processes}, \bibinfo{publisher}{Wiley}, \bibinfo{address}{New
  York}, \bibinfo{year}{1967}.

\bibitem[{Diener and Poston(1981)}]{Poston1981}
\bibinfo{author}{M.~Diener}, \bibinfo{author}{T.~Poston}, \bibinfo{title}{Chaos
  and Order in Nature: Proc. Int. Symp. Energetics}, in:
  \bibinfo{editor}{H.~Haken} (Ed.), \bibinfo{booktitle}{Chaos and Order in
  Nature: Proc. Int. Symp. Energetics}, \bibinfo{publisher}{Springer},
  \bibinfo{year}{1981}.

\bibitem[{Arfken(1985)}]{Arfken}
\bibinfo{author}{G.~Arfken}, \bibinfo{title}{Mathematical Methods For
  Physicists}, \bibinfo{publisher}{Academic Press}, \bibinfo{edition}{third}
  edn., \bibinfo{year}{1985}.

\bibitem[{Savary et~al.(2012)Savary, Ross, Gaulin, Ruff, and
  Balents}]{PhysRevLett.109.167201}
\bibinfo{author}{L.~Savary}, \bibinfo{author}{K.~A. Ross},
  \bibinfo{author}{B.~D. Gaulin}, \bibinfo{author}{J.~P.~C. Ruff},
  \bibinfo{author}{L.~Balents}, \bibinfo{title}{Order by Quantum Disorder in
  ${\mathrm{Er}}_{2}{\mathrm{Ti}}_{2}{\mathbf{O}}_{7}$},
  \bibinfo{journal}{Phys. Rev. Lett.} \bibinfo{volume}{109}
  (\bibinfo{year}{2012}) \bibinfo{pages}{167201},
  \doi{\bibinfo{doi}{10.1103/PhysRevLett.109.167201}}.

\bibitem[{Barnett et~al.(2012)Barnett, Powell, Gra\ss{}, Lewenstein, and
  Das~Sarma}]{Barnett2012}
\bibinfo{author}{R.~Barnett}, \bibinfo{author}{S.~Powell},
  \bibinfo{author}{T.~Gra\ss{}}, \bibinfo{author}{M.~Lewenstein},
  \bibinfo{author}{S.~Das~Sarma}, \bibinfo{title}{Order by disorder in
  spin-orbit-coupled Bose-Einstein condensates}, \bibinfo{journal}{Phys. Rev.
  A} \bibinfo{volume}{85} (\bibinfo{year}{2012}) \bibinfo{pages}{023615},
  \doi{\bibinfo{doi}{10.1103/PhysRevA.85.023615}}.

\bibitem[{Villain et~al.(1980)Villain, Bidaux, Carton, and Conte}]{Villain1980}
\bibinfo{author}{J.~Villain}, \bibinfo{author}{R.~Bidaux},
  \bibinfo{author}{J.~P. Carton}, \bibinfo{author}{R.~Conte},
  \bibinfo{title}{Order as an Effect of Dirorder}, \bibinfo{journal}{J. de
  Physique} \bibinfo{volume}{41} (\bibinfo{year}{1980})
  \bibinfo{pages}{1263--1272}.

\bibitem[{Oitmaa et~al.(2013)Oitmaa, Singh, Javanparast, Day, Bagheri, and
  Gingras}]{PhysRevB.88.220404}
\bibinfo{author}{J.~Oitmaa}, \bibinfo{author}{R.~R.~P. Singh},
  \bibinfo{author}{B.~Javanparast}, \bibinfo{author}{A.~G.~R. Day},
  \bibinfo{author}{B.~V. Bagheri}, \bibinfo{author}{M.~J.~P. Gingras},
  \bibinfo{title}{Phase transition and thermal order-by-disorder in the
  pyrochlore antiferromagnet Er${}_{2}$Ti${}_{2}$O${}_{7}$: A high-temperature
  series expansion study}, \bibinfo{journal}{Phys. Rev. B} \bibinfo{volume}{88}
  (\bibinfo{year}{2013}) \bibinfo{pages}{220404},
  \doi{\bibinfo{doi}{10.1103/PhysRevB.88.220404}}.

\bibitem[{Nicolis and Prigogine(1977)}]{Nicolis1977}
\bibinfo{author}{G.~Nicolis}, \bibinfo{author}{I.~Prigogine},
  \bibinfo{title}{Self-Organization in Nonequilibrium Systems: From Dissipative
  Structures to Order through Fluctuations}, \bibinfo{publisher}{Wiley},
  \bibinfo{address}{New York}, \bibinfo{year}{1977}.

\bibitem[{de~Groot(1966)}]{degroot}
\bibinfo{author}{S.~de~Groot}, \bibinfo{title}{Thermodynamics of Irreversible
  Processes}, \bibinfo{publisher}{North Holland}, \bibinfo{address}{Amsterdam},
  \bibinfo{year}{1966}.

\bibitem[{Zangwill(1988)}]{Zangwill}
\bibinfo{author}{A.~Zangwill}, \bibinfo{title}{Physics at Surfaces},
  \bibinfo{publisher}{Cambridge University Press}, \bibinfo{address}{Cambridge,
  England}, \bibinfo{year}{1988}.

\bibitem[{{CRC Handbook}(2007)}]{noauthororeditor2007handbook}
\bibinfo{author}{{CRC Handbook}}, \bibinfo{title}{CRC Handbook of Chemistry and
  Physics, 88th Edition}, \bibinfo{publisher}{CRC Press}, \bibinfo{edition}{88}
  edn., ISBN \bibinfo{isbn}{0849304881}, \bibinfo{year}{2007}.

\bibitem[{Meyer and Reif(1961)}]{Reif1961}
\bibinfo{author}{L.~Meyer}, \bibinfo{author}{F.~Reif}, \bibinfo{title}{Ion
  Motion in Superfluid Liquid Helium Under Pressure}, \bibinfo{journal}{Phys.
  Rev.} \bibinfo{volume}{123} (\bibinfo{year}{1961}) \bibinfo{pages}{727}.

\bibitem[{Harrison(1970)}]{HarrisonSolid}
\bibinfo{author}{W.~A. Harrison}, \bibinfo{title}{Solid State Theory},
  \bibinfo{publisher}{Courier Dover}, \bibinfo{address}{New York},
  \bibinfo{year}{1970}.

\bibitem[{Ashcroft and Mermin(1976)}]{AM}
\bibinfo{author}{N.~W. Ashcroft}, \bibinfo{author}{D.~N. Mermin},
  \bibinfo{title}{{Solid State Physics}}, \bibinfo{publisher}{Thomson
  Learning}, \bibinfo{address}{Toronto}, \bibinfo{edition}{1} edn., ISBN
  \bibinfo{isbn}{0030839939}, \bibinfo{year}{1976}.

\bibitem[{Sze(1981)}]{Sze}
\bibinfo{author}{S.~Sze}, \bibinfo{title}{Physics of Semiconductor Devices},
  \bibinfo{publisher}{Wiley}, \bibinfo{edition}{2nd} edn.,
  \bibinfo{year}{1981}.

\bibitem[{Ninomiya et~al.(2009)Ninomiya, Ichiki, Yamada, Nakata, Seki, Aoki,
  and Matsuo}]{RCM:RCM4046}
\bibinfo{author}{S.~Ninomiya}, \bibinfo{author}{K.~Ichiki},
  \bibinfo{author}{H.~Yamada}, \bibinfo{author}{Y.~Nakata},
  \bibinfo{author}{T.~Seki}, \bibinfo{author}{T.~Aoki},
  \bibinfo{author}{J.~Matsuo}, \bibinfo{title}{Precise and fast secondary ion
  mass spectrometry depth profiling of polymer materials with large Ar cluster
  ion beams}, \bibinfo{journal}{Rapid Communications in Mass Spectrometry}
  \bibinfo{volume}{23}~(\bibinfo{number}{11}) (\bibinfo{year}{2009})
  \bibinfo{pages}{1601--1606}, ISSN \bibinfo{issn}{1097-0231},
  \doi{\bibinfo{doi}{10.1002/rcm.4046}},
  \urlprefix\url{http://dx.doi.org/10.1002/rcm.4046}.

\bibitem[{Stroscio and Kaiser(1993)}]{STMbook}
\bibinfo{author}{J.~Stroscio}, \bibinfo{author}{W.~Kaiser},
  \bibinfo{title}{Scanning Tunneling Microscopy}, \bibinfo{publisher}{Academic
  Press}, \bibinfo{address}{NY}, \bibinfo{year}{1993}.

\bibitem[{Walkup et~al.(1993)Walkup, Newns, and Avouris}]{Avouris93}
\bibinfo{author}{R.~E. Walkup}, \bibinfo{author}{D.~M. Newns},
  \bibinfo{author}{P.~Avouris}, \bibinfo{title}{Role of multiple inelastic
  transitions in atom transfer with the scanning tunneling microscope},
  \bibinfo{journal}{Phys. Rev. B} \bibinfo{volume}{48} (\bibinfo{year}{1993})
  \bibinfo{pages}{1858}.

\bibitem[{Estrup et~al.(1986)Estrup, Greene, Cardillo, and
  Tully}]{EstrupTully1986}
\bibinfo{author}{P.~J. Estrup}, \bibinfo{author}{E.~F. Greene},
  \bibinfo{author}{M.~J. Cardillo}, \bibinfo{author}{J.~C. Tully},
  \bibinfo{title}{Influence of Surface Phase Transitions on Desorption
  Kinetics: The Compensation Effect}, \bibinfo{journal}{J. Phys. Chem.}
  \bibinfo{volume}{90} (\bibinfo{year}{1986}) \bibinfo{pages}{4099--4104}.

\bibitem[{Brown et~al.(1998)Brown, Moffatt, and Wolkow}]{Wolkow1998}
\bibinfo{author}{D.~Brown}, \bibinfo{author}{D.~Moffatt},
  \bibinfo{author}{R.~Wolkow}, \bibinfo{title}{Isolation of an Intrinsic
  Precursor to Molecular Chemisorption}, \bibinfo{journal}{Sci.}
  \bibinfo{volume}{279} (\bibinfo{year}{1998}) \bibinfo{pages}{542}.

\bibitem[{Villarrubia and Boland(1989)}]{Boland1989}
\bibinfo{author}{J.~S. Villarrubia}, \bibinfo{author}{J.~J. Boland},
  \bibinfo{title}{Scanning-tunneling-microscopy study of the Si(111)-7x7
  rest-atom layer following adatom removal by reaction with Cl},
  \bibinfo{journal}{Phys. Rev. Lett.} \bibinfo{volume}{63}
  (\bibinfo{year}{1989}) \bibinfo{pages}{306--309},
  \doi{\bibinfo{doi}{10.1103/PhysRevLett.63.306}}.

\bibitem[{Liu et~al.(2009)Liu, Horn, Maraghechi, and Patitsas}]{Patitsas2009}
\bibinfo{author}{W.~Liu}, \bibinfo{author}{S.~Horn},
  \bibinfo{author}{P.~Maraghechi}, \bibinfo{author}{S.~Patitsas},
  \bibinfo{journal}{J. Vac. Sci. Tech. B} \bibinfo{volume}{27}
  (\bibinfo{year}{2009}) \bibinfo{pages}{895}.

\bibitem[{Lyo and Avouris(1991)}]{Avouris1991}
\bibinfo{author}{I.~W. Lyo}, \bibinfo{author}{P.~Avouris},
  \bibinfo{title}{Field-induced nanometer-to atomic scale manipulation of
  silicon surfaces with the STM}, \bibinfo{journal}{Science}
  \bibinfo{volume}{253} (\bibinfo{year}{1991}) \bibinfo{pages}{173}.

\bibitem[{Zeppenfeld et~al.(1992)Zeppenfeld, Lutz, and Eigler}]{Eigler92}
\bibinfo{author}{P.~Zeppenfeld}, \bibinfo{author}{C.~P. Lutz},
  \bibinfo{author}{D.~M. Eigler}, \bibinfo{title}{Manipulating Atoms and
  Molecules with a Scanning Tunneling Microscope},
  \bibinfo{journal}{Ultramicroscopy} \bibinfo{volume}{42-44}
  (\bibinfo{year}{1992}) \bibinfo{pages}{128--133}.

\bibitem[{Nabighian and Zhu(2000)}]{Nabighian2000}
\bibinfo{author}{E.~Nabighian}, \bibinfo{author}{X.~D. Zhu},
  \bibinfo{title}{Diffusion of Xe on Ni(111)}, \bibinfo{journal}{Chem. Phys.
  Lett.} \bibinfo{volume}{316}~(\bibinfo{number}{3-4}) (\bibinfo{year}{2000})
  \bibinfo{pages}{177 -- 180}, ISSN \bibinfo{issn}{0009-2614},
  \doi{\bibinfo{doi}{10.1016/S0009-2614(99)01292-0}}.

\bibitem[{Bartels et~al.(1997)Bartels, Meyer, and Rieder}]{Meyer97}
\bibinfo{author}{L.~Bartels}, \bibinfo{author}{G.~Meyer},
  \bibinfo{author}{K.~H. Rieder}, \bibinfo{title}{Basic Steps of Lateral
  Manipulation of Single Atoms and Diatomic Clusters with a Scanning Tunneling
  Microscope Tip}, \bibinfo{journal}{Phys. Rev. Lett.}
  \bibinfo{volume}{79}~(\bibinfo{number}{4}) (\bibinfo{year}{1997})
  \bibinfo{pages}{697--700}, \doi{\bibinfo{doi}{10.1103/PhysRevLett.79.697}}.

\bibitem[{Whitman et~al.(1991)Whitman, Stroscio, Dragoset, and
  Celotta}]{StroscioCelotta91}
\bibinfo{author}{L.~J. Whitman}, \bibinfo{author}{J.~A. Stroscio},
  \bibinfo{author}{R.~A. Dragoset}, \bibinfo{author}{R.~J. Celotta},
  \bibinfo{title}{Manipulation of Adsorbed Atoms and Creation of New Structures
  on Room-Temperature Surfaces with a Scanning Tunneling Microscope},
  \bibinfo{journal}{Science} \bibinfo{volume}{251} (\bibinfo{year}{1991})
  \bibinfo{pages}{1206--1210}.

\bibitem[{Martsinovich and Kantorovich(2008)}]{Kantorovich2008}
\bibinfo{author}{N.~Martsinovich}, \bibinfo{author}{L.~Kantorovich},
  \bibinfo{title}{Pulling the $C_{60}$ molecule on a Si(001) surface with an
  STM tip: A theoretical study}, \bibinfo{journal}{Phys. Rev. B}
  \bibinfo{volume}{77}~(\bibinfo{number}{11}) (\bibinfo{year}{2008})
  \bibinfo{pages}{115429}, \doi{\bibinfo{doi}{10.1103/PhysRevB.77.115429}}.

\bibitem[{Ness and Fisher(1999)}]{Fisher1999}
\bibinfo{author}{H.~Ness}, \bibinfo{author}{A.~J. Fisher},
  \bibinfo{title}{Quantum Inelastic Conductance Through Molecular Wires},
  \bibinfo{journal}{Physical Review Letters} \bibinfo{volume}{83}
  (\bibinfo{year}{1999}) \bibinfo{pages}{452}.

\bibitem[{Lopinski et~al.(1998)Lopinski, Fortier, Moffatt, and
  Wolkow}]{Lopinski1998}
\bibinfo{author}{G.~Lopinski}, \bibinfo{author}{T.~Fortier},
  \bibinfo{author}{D.~Moffatt}, \bibinfo{author}{R.~Wolkow},
  \bibinfo{journal}{J. Vac. Sci. Tech. A} \bibinfo{volume}{16}
  (\bibinfo{year}{1998}) \bibinfo{pages}{1037}.

\bibitem[{Landauer(1957)}]{Landauer1957}
\bibinfo{author}{R.~Landauer}, \bibinfo{title}{Spatial Variation of Currents
  and Fields Due to Localized Scatterers in Metallic Conduction},
  \bibinfo{journal}{J. IBM Res. Dev.} \bibinfo{volume}{1}
  (\bibinfo{year}{1957}) \bibinfo{pages}{223--231}.

\bibitem[{McLennan(1974)}]{McLennan1974}
\bibinfo{author}{J.~A. McLennan}, \bibinfo{title}{Onsager's theorem and
  higher-order hydrodynamic equations}, \bibinfo{journal}{Phys. Rev. A}
  \bibinfo{volume}{10} (\bibinfo{year}{1974}) \bibinfo{pages}{1272--1276},
  \doi{\bibinfo{doi}{10.1103/PhysRevA.10.1272}}.

\bibitem[{Dufty and Rub\'\i(1987)}]{Dufty1987}
\bibinfo{author}{J.~W. Dufty}, \bibinfo{author}{J.~M. Rub\'\i},
  \bibinfo{title}{Generalized Onsager symmetry}, \bibinfo{journal}{Phys. Rev.
  A} \bibinfo{volume}{36} (\bibinfo{year}{1987}) \bibinfo{pages}{222--225},
  \doi{\bibinfo{doi}{10.1103/PhysRevA.36.222}}.

\bibitem[{McLennan(1961)}]{McLennan1961}
\bibinfo{author}{J.~A. McLennan}, \bibinfo{title}{Nonlinear Effects in
  Transport Theory}, \bibinfo{journal}{Phys. Fluids} \bibinfo{volume}{4}
  (\bibinfo{year}{1961}) \bibinfo{pages}{1319}.

\bibitem[{Dufty and McLennan(1968)}]{McLennan1968}
\bibinfo{author}{J.~W. Dufty}, \bibinfo{author}{J.~A. McLennan},
  \bibinfo{title}{Correlation Functions as Hydrodynamic Green's Functions},
  \bibinfo{journal}{Phys. Rev.} \bibinfo{volume}{172} (\bibinfo{year}{1968})
  \bibinfo{pages}{176--181}, \doi{\bibinfo{doi}{10.1103/PhysRev.172.176}}.

\bibitem[{Dufty(1968)}]{Dufty1968}
\bibinfo{author}{J.~W. Dufty}, \bibinfo{title}{Hydrodynamics and
  Time-Correlation Functions}, \bibinfo{journal}{Phys. Rev.}
  \bibinfo{volume}{176} (\bibinfo{year}{1968}) \bibinfo{pages}{398--409},
  \doi{\bibinfo{doi}{10.1103/PhysRev.176.398}}.

\bibitem[{Wang and Heinz(2002)}]{Heinz2002}
\bibinfo{author}{E.~Wang}, \bibinfo{author}{U.~Heinz},
  \bibinfo{title}{Generalized fluctuation-dissipation theorem for nonlinear
  response functions}, \bibinfo{journal}{Phys. Rev. D} \bibinfo{volume}{66}
  (\bibinfo{year}{2002}) \bibinfo{pages}{025008}.

\bibitem[{Tersoff and Hamann(1985)}]{Tersoff1985}
\bibinfo{author}{J.~Tersoff}, \bibinfo{author}{D.~R. Hamann},
  \bibinfo{journal}{Phys. Rev. B} \bibinfo{volume}{31} (\bibinfo{year}{1985})
  \bibinfo{pages}{805}.

\bibitem[{Meyer et~al.(2003)Meyer, Hug, and Bennewitz}]{MeyerAFMbook}
\bibinfo{author}{E.~Meyer}, \bibinfo{author}{H.~Hug},
  \bibinfo{author}{R.~Bennewitz}, \bibinfo{title}{Scanning Probe Microscopy:
  The Lab on a Tip}, \bibinfo{publisher}{Springer}, \bibinfo{address}{New
  York}, \bibinfo{year}{2003}.

\end{thebibliography}
 \bibliographystyle{elsarticle-num-names.bst}

\end{document}